\documentclass[fleqn,usenatbib]{mnras}

\usepackage[T1]{fontenc}
\usepackage{ae,aecompl}

\usepackage{graphicx}	
\usepackage{amsmath}	
\usepackage{amssymb}	

\usepackage{textgreek}
\usepackage[flushleft]{threeparttable}
\usepackage{enumitem}
\usepackage{float}
\usepackage{caption}
\usepackage{pdflscape}
\usepackage{subcaption}
\usepackage{cleveref}

\usepackage{soul}
\usepackage{multirow}
\DeclareMathAlphabet{\mathcal}{OMS}{cmsy}{m}{n}
\usepackage{newtxtext,newtxmath}

\title[WASP-19b ESPRESSO Transmission Spectroscopy]{A Spectral Survey of WASP-19b with ESPRESSO}

\author[E. Sedaghati et al.]{
Elyar Sedaghati$^{1,2,\dag,}$\thanks{E-mail: esedagha@eso.org},
Ryan J. MacDonald$^{3}$,
N\'uria Casasayas-Barris$^{4}$,
\newauthor
H. Jens Hoeijmakers$^{5}$,
Henri M. J. Boffin$^{6}$,
Florian Rodler$^{1}$,
Rafael Brahm$^{7,8}$,
\newauthor
Mat\'ias Jones$^{1,9,\dag}$,
Alejandro S\'anchez-L\'opez$^{4}$,
Ilaria Carleo$^{10}$,
\newauthor
Pedro Figueira$^{1,11}$,
Andrea Mehner$^{1}$,
Manuel L\'opez-Puertas$^{2}$
\\
$^1$European Southern Observatory, Alonso de C\'ordova 3107, Santiago, Chile\\
$^2$Instituto de Astrof\'isica de Andaluc\'ia (IAA-CSIC), Glorieta de la Astronom\'ia s/n, 18008 Granada, Spain\\
$^3$Department of Astronomy and Carl Sagan Institute, Cornell University, 122 Sciences Drive, Ithaca, NY 14853, USA\\
$^4$Leiden Observatory, Leiden University, Postbus 9513, 2300 RA, Leiden, The Netherlands\\
$^5$Department of Astronomy and Theoretical Physics, Lund University, S\"olvegatan 27, 223 62 Lund, Sweden\\
$^6$European Southern Observatory, Karl-Schwarzschild-Stra{\ss}e 2, 85748 Garching bei M\"unchen, Germany\\
$^7$Facultad de Ingenier\'ia y Ciencias, Universidad Adolfo Ib\'a\~nez, Diagonal Las Torres 2640, Pe\~nalol\'en, Chile\\
$^8$Millennium Institute for Astrophysics, Chile\\
$^9$Instituto de Astronom\'ia, Universidad Cat\'olica del Norte, Angamos 0610, 1270709, Antofagasta, Chile \\
$^{10}$Astronomy Department and Van Vleck Observatory, Wesleyan University, Middletown, CT 06459, USA\\
$^{11}$Instituto de Astrof\'{i}sica e Ci\^{e}ncias do Espa\c{c}o, Universidade do Porto,CAUP, Rua das Estrelas, 4150-762 Porto, Portugal\\
$^{\dag}$ ESO Fellow
}

\date{Accepted 2021 April 20. Received 2021 April 20; in original form 2021 February 4.}

\pubyear{2021}

\begin{document}
\label{firstpage}
\pagerange{\pageref{firstpage}--\pageref{lastpage}}
\maketitle

\begin{abstract}
High resolution precision spectroscopy provides a multitude of robust techniques for probing exoplanetary atmospheres. We present multiple VLT/ESPRESSO transit observations of the hot-Jupiter exoplanet WASP-19b with previously published but disputed atmospheric features from low resolution studies. Through spectral synthesis and modeling of the Rossiter-McLaughlin (RM) effect we calculate stellar, orbital and physical parameters for the system. From narrow-band spectroscopy we do not detect any of H\,{\scriptsize I}, Fe\,{\scriptsize I}, Mg\,{\scriptsize I}, Ca\,{\scriptsize I}, Na\,{\scriptsize I} and K\,{\scriptsize I} neutral species, placing upper limits on their line contrasts. Through cross correlation analyses with 
atmospheric models, we do not detect Fe\,{\scriptsize I} and place a 3$\sigma$ upper limit of $\log\,(X_{\textrm{Fe}}/X_\odot) \approx -1.83\,\pm\,0.11$ on its mass fraction, from injection and retrieval. We show the inability to detect the presence of H$_2$O for known abundances, owing to lack of strong absorption bands, as well as relatively low S/N ratio. We detect a barely significant peak (3.02\,$\pm$\,0.15\,$\sigma$) in the cross correlation map for TiO, consistent with the sub-solar abundance previously reported. This is merely a hint for the presence of TiO and does \textit{not} constitute a confirmation. However, we do confirm the presence of previously observed enhanced scattering towards blue wavelengths, through chromatic RM measurements, pointing to a hazy atmosphere. We finally present a reanalysis of low resolution transmission spectra of this exoplanet, concluding that unocculted starspots alone cannot explain previously detected features. Our reanalysis of the FORS2 spectra of WASP-19b finds a $\sim$\,100$\times$ sub-solar TiO abundance, precisely constrained to $\log\,X_{\textrm{TiO}} \approx -7.52 \pm 0.38$, consistent with the TiO hint from ESPRESSO. We present plausible paths to reconciliation with other seemingly contradicting results.
\end{abstract}





\section{Introduction}
Detections of ionic, atomic and molecular species in exoplanetary atmospheres serve as a unique and strong diagnostic of those chemical and dynamical processes driving their formation and evolution. Their detection and abundance measurements could act as indicators of planetary formation scenarios and reveal connections to the primordial protoplanetary disk and the host star \citep{Williams2011,Mordasini2016,Madhusudhan2017}. Furthermore, discoveries of atmospheric chemical species allow us to better understand various thermodynamical processes and chemistry, winds in the upper atmosphere \citep{Goodman2009,Snellen2010,Brogi2016,Madhusudhan2016,Wyttenbach2020}, and to probe planetary interiors and various bulk properties through their abundances \citep{Kite2016,Thorngren2019,Madhusudhan2020}. A whole host of ions, atoms and molecules have been detected through a variety of, often complementary, techniques, such as differential spectrophotometry using low-to-mid resolution spectroscopy \citep[e.g.][]{Gibson2012,Gibson2017,Deming2013,Kreidberg2014,Kirk2016,Nortmann2016}, and high resolution spectroscopic techniques \citep[e.g.][]{Redfield2008,Snellen2008,Rodler2012,Birkby2013,Hoeijmakers2015,Hoeijmakers2018,Hoeijmakers2020,Brogi2016, Birkby2017,Zak2019,Ehrenreich2020}.  To date, ionic species such as Fe\,{\scriptsize II} and Ti\,{\scriptsize II} \citep{Hoeijmakers2019}, atomic absorption from Na, K, H$\alpha$ and He \citep[e.g.][]{Redfield2008,Sedaghati2016,Casasayas2017,Spake2018,Chen2020,Seidel2020}, and molecules such as H$_2$O, CH$_4$ and CO \citep[e.g.][]{Konopacky2013,Brogi2014,Fraine2014,Barman2015,Sing2016}, have been detected through the aforementioned techniques. Needless to say that this list of detected constituents is by no means exhaustive, nor that of methods employed to detect exoplanetary atmospheres. For instance, high resolution imaging instruments such as SPHERE \citep{Beuzit2019} and GRAVITY \citep{Gravity2017}, both at the VLT (ESO's Very Large Telescope), through combination with low dispersion spectroscopy, have facilitated direct measurements of exoplanetary atmospheres \citep{Samland2017,Nowak2020}.

A complete inventory of heavy element enrichment in an exoplanetary atmosphere informs us of planetary formation mechanisms in a way that most direct methods are unable to do. While core accretion implies that giant exoplanets are expected to be heavily enriched in metals \citep{Mordasini2009}, tidal downsizing \citep{Nayakshin2010} and other flavours of gravitational instability suggest that planets should have the same fraction of heavy elements as their host star. Moreover, with the discovery of the radius gap by \citet{Fulton2017} and predictions for the Fe content \citep{Owen2018}, significant recent progress has been made in associating atmospheric abundances with those formation processes.

In this study, we employ high dispersion spectroscopy ($\mathcal{R} \gtrsim 10^5$) to probe an exoplanetary atmosphere to constrain its chemical inventory, as well as possible dynamical characteristics. This versatile technique resolves individual spectral lines, allowing one to probe the exoplanetary atmosphere through a variety of approaches. First, one can search for excess absorption in the cores of strong singular transition lines emanating from the atmosphere of a transiting planet via differential in-transit and outside-transit residual spectrum analysis \citep[e.g.][]{Wyttenbach2017,Casasayas2019,Chen2020}. Secondly, phase-resolved atmospheric absorption, reflection or emission, from both transiting and non-transiting planets, can be mapped out through cross correlation with model templates \citep[e.g.][]{Brogi2014,Hoeijmakers2015,Hoeijmakers2018,Hoeijmakers2019,Hoeijmakers2020,Pino2020,Yan2020}. Finally, chromatic Rossiter-McLaughlin \citep[CRM;][]{Snellen2004,Dreizler2009} measurements can probe the atmosphere of a transiting exoplanet in a similar manner to spectrophotometric studies by measuring the planetary radius as a function of wavelength \citep[e.g.][]{DiGloria2015,Borsa2016,Borsa2020,Boldt2020,Oshagh2020,Santos2020}. This final approach comes at no extra observational cost and allows to probe different atmospheric layers from the same set of observations.

We present observations of the ultra-short period hot-Jupiter exoplanet WASP-19b \citep{Hebb2010} on a transiting orbit around a G8V, 12.3\,V magnitude star. It orbits its host star with a 0.78884\,day period and has a mass and radius of 1.168\,$\pm$\,0.023\,M$_{\textrm{\textit{jup}}}$ and 1.386\,$\pm$\,0.032\,R$_{\textrm{\textit{jup}}}$ \citep{Hellier2011}. Transmission and emission spectroscopy indicate a temperature of $\sim$\,2200\,K \citep{Sedaghati2017,Wong2020}, putting it on the boundary of hot to ultra-hot Jupiter exoplanets \citep{Parmentier2018}.

WASP-19b's atmosphere has been extensively studied at low spectral resolution, albeit with somewhat discrepant interpretations \citep{Bean2013,Lendl2013,Mancini2013,Mandell2013,Tregloan-Reed2013}. \citet{Huitson2013} used the Hubble Space Telescope (HST) to obtain a visible to near-IR transmission spectrum of WASP-19b, reporting a significant (4$\sigma$) detection of H$_2$O from their WFC3 spectra and possibly ruling out solar abundance TiO and alkali line features from their STIS observations (at $\sim$\,2.8$\sigma$ confidence). However, they determined that their light curves obtained from the visible channel were heavily modulated by the crossing of stellar active regions (spots), rendering the visible wavelength transmission spectrum challenging to precisely measure. \citet{Wong2016}, using Spitzer 3.6 and 4.5 $\mu$m channels, reported emission from the day-side of WASP-19b consistent with no thermal inversion and a moderately efficient day-night atmospheric circulation. Additionally, their results suggest the possible presence of a super-rotating equatorial jet, while their phase curves hinted at high-altitude silicate clouds in the night-side and/or a high atmospheric metallicity. Building on those results, \citet{Wong2020} measured a strong atmospheric modulation signal from TESS \citep{Ricker2015} secondary eclipse observations, reported no offset of the maximum brightness point from the sub-stellar point from the phase-curve analysis and derived a day-side temperature 2240\,$\pm$\,40\,K from the retrieval of the emission spectrum. As part of a large atmospheric survey using the HST, \citet{Sing2016} concluded that the resolution of the obtained spectrum in the optical range was not high enough for any detection of individual alkali or molecular species. The only conclusion drawn from the optical spectrum was an enhanced planetary radius toward near-UV wavelengths, attributed to Rayleigh scattering in the upper atmosphere. 

\citet{Sedaghati2017} presented the detection of TiO (7.7\,$\sigma$), a steep optical slope, as well as the confirmation of H$_2$O, from a retrieval analysis of multi-grism FORS2/VLT spectrophotometric transmission spectra ($\mathcal{R} \sim 73$) of WASP-19b. This pointed to an atmosphere with high altitude hazes and sub-solar abundances of metal oxides. Following this study, \citet{Espinoza2019} produced another optical transmission spectrum ($\mathcal{R} \sim 40$) of WASP-19b using the IMACS/Magellan instrument, also from multi-epoch observations. Their combined spectrum did not confirm the presence of either TiO or the near-UV scattering slope, though evidence of TiO and a slope in individual epochs were attributed to low contrast stellar active regions unocculted by the transiting planet. They suggested that WASP-19b's atmosphere contains high altitude clouds, masking any optical features, and that those detected from the FORS2 spectrum are likely due to stellar contamination (i.e., the transit light source effect -- see  \cite{Rackham2017}). These contrasting conclusions provide the main motivation for this study. High resolution observations can readily distinguish stellar contamination from signals of planetary origin, given the substantial differences between the stellar rotation and the planetary orbital velocity. Such differences lead to two separate signals occupying different locations in the cross-correlation velocity space, as detailed further in section \ref{section:CCF analysis}.

In section \ref{sec:Obs} we present the ESPRESSO observations and data reduction, as well as our methodology for mitigating the systematic noise in the raw frames attributed to the readout electronics and optical path inhomogeneities. In section \ref{sec:Methods} we demonstrate our data analysis methods including all steps necessary for probing the atmosphere through transmission spectroscopy with narrow-band and cross correlation techniques. In section \ref{sec:Results} we show the resulting atmospheric inferences from our WASP-19b ESPRESSO observations. Section \ref{sec:Discussion} presents additional analysis of the exoplanetary atmosphere through chromatic Rossiter-McLaughlin measurements, as well as a reanalyses of low resolution transmission spectra of WASP-19b through atmospheric retrievals including unocculted stellar heterogeneities. Finally, in section \ref{sec:Conclusions} we conclude the study by summarizing our results.

\section{Observations \& data reduction}
\label{sec:Obs}
We observed four primary transits of WASP-19b between 15-01-2019 and 12-01-2020 (see table \ref{tab:Observations} for a full summary), with the ESPRESSO spectrograph at the VLT \citep{Pepe2010,Pepe2014,Pepe2021}. ESPRESSO is the extremely stable, fiber-fed high resolution, cross-dispersed echelle spectrograph installed at the Incoherent Combined Coud\'e Focus (ICCF) of the VLT, capable of light injection from any of the UT's at a time (1-UT mode), as well as all 4 telescopes simultaneously (4-UT mode). In all configurations the spectrograph is illuminated by two fibers; one used on target and the other for the purpose of sky background observation or light from a reference source. It has two cameras for focusing and recording the orders of the echellogram, each of which is separated into two slices by the anamorphic pupil slicing unit \citep[APSU;][]{Riva2014,Oggioni2016} in order to keep the size of the grating reasonably small.

All observations were taken under the ESO programme ID 0102.C-0311 (PI: Sedaghati). These data sets are referred to as DS1 to DS4 in chronological order in this paper. All observations were performed in High Resolution 1-UT mode, which utilises the fiber core equivalent to 1$^{\prime\prime}$ on the sky. This translates to spectra at $\mathcal{R} \sim 140~000$, covering the approximate wavelength range of $3770 - 7900$\,\AA. DS1--DS3 were taken in unbinned (1$\times$1) readout mode in order to reduce CCD readout time, and subsequently increase the duty cycle. However, after the initial analysis, it became clear that the associated 500\,kpx/sec readout speed introduces a correlated noise at the bias level, much larger than the binned (2$\times$1) readout which reads data at a much lower frequency (100\,kpx/sec). This becomes particularly problematic at the low signal-to-noise (S/N) regime. Therefore, for DS4 the binned readout mode was adopted in order to reduce somewhat the electronic noise. It must however be noted that there still remain some detector artefacts, which are discussed in the following subsection. 

We adopted the unorthodox approach of different in- and out-of-transit exposure times for all sets of observations, a short summary of which is presented in table \ref{tab:Observations}, with full details given in \crefrange{tab:RV1}{tab:RV4}. This choice is dictated by the magnitude of the star and the short duration of the transit. Namely, larger exposure times are chosen to maximize the duty cycle of the observations out of transit, while shorter exposure times in transit ensure that any potential planetary signal is not smeared due to large changes in the planet's radial velocity throughout an exposure.

For all observations, fiber B was placed on the sky in order to monitor any potential contamination from the moon, although all epochs were chosen in such a way to maximise lunar separation. The four data sets were taken under varying sky transparency conditions (see Table \ref{tab:Observations}). It must also be mentioned that DS4 was taken after the technical mission of July 2019 that upgraded the ESPRESSO fibers, which resulted in 
substantial gains in transmission \citep[$\leq$\,40\%;][]{Pepe2021}. This fact is clearly reflected in the increase in the S/N of spectra, as well as precision in the obtained individual results, considering the discrepant observing conditions.

\begin{table*}
    \caption{Observational details of all transits with ESPRESSO.}
    \begin{tabular}{lccccccccc}
    \hline
     Data & Tel.  & Date & Time & N$_{\text{obs}}$& Exp. time      & Airmass & S/N @550nm$^a$ & Sky        & Seeing \\
     ID   &       &      & [UT]   & (in-transit)    & (in-transit) [s]&        & (in-transit) & transmission & [$^{\prime\prime}$]\\
     \hline \hline
     DS1 & UT3 & 15-01-19 & 01:57 $\rightarrow$ 05:32 & 22 (12) & > 760 (380) & 2.17 $\rightarrow$ 1.13 & 12 -- 26~ (9) & CLR $\rightarrow$ PHO & 0.44 -- 0.86 \\
     DS2 & UT3 & 04-03-19 & 04:34 $\rightarrow$ 08:21 & 20 (12) & > 670 (410) & 1.09 $\rightarrow$ 1.86 & 10 -- 21~ (8) & CLR & 1.30 -- 2.25\\
     DS3 & UT3 & 23-03-19 & 02:53 $\rightarrow$ 06:39 & 22 (12) & > 720 (380) & 1.07 $\rightarrow$ 1.66 & 25 -- 28 (15) & CLR & 0.30 -- 0.60\\
     DS4 & UT1 & 12-01-20 & 03:35 $\rightarrow$ 07:25 & 22 (11) & > 750 (380) & 1.49 $\rightarrow$ 1.07 & 29 -- 35 (25) & THN & 0.48 -- 0.84\\
     \hline
    \multicolumn{10}{l}{$^a$ Quadrature sum of the two S/N values for the relevant slices. The in-transit spectra have lower S/N values due to shorter exposure}\\
    \multicolumn{10}{l}{~~~times. This is necessary not to smear any potential atmospheric signal due to the change in planetary RV during the exposure.}
    \end{tabular}
  \label{tab:Observations}
\end{table*}

\subsection{Correlated noise correction}
\label{sec:electronic noise}
\begin{figure}
	\includegraphics[width=\columnwidth]{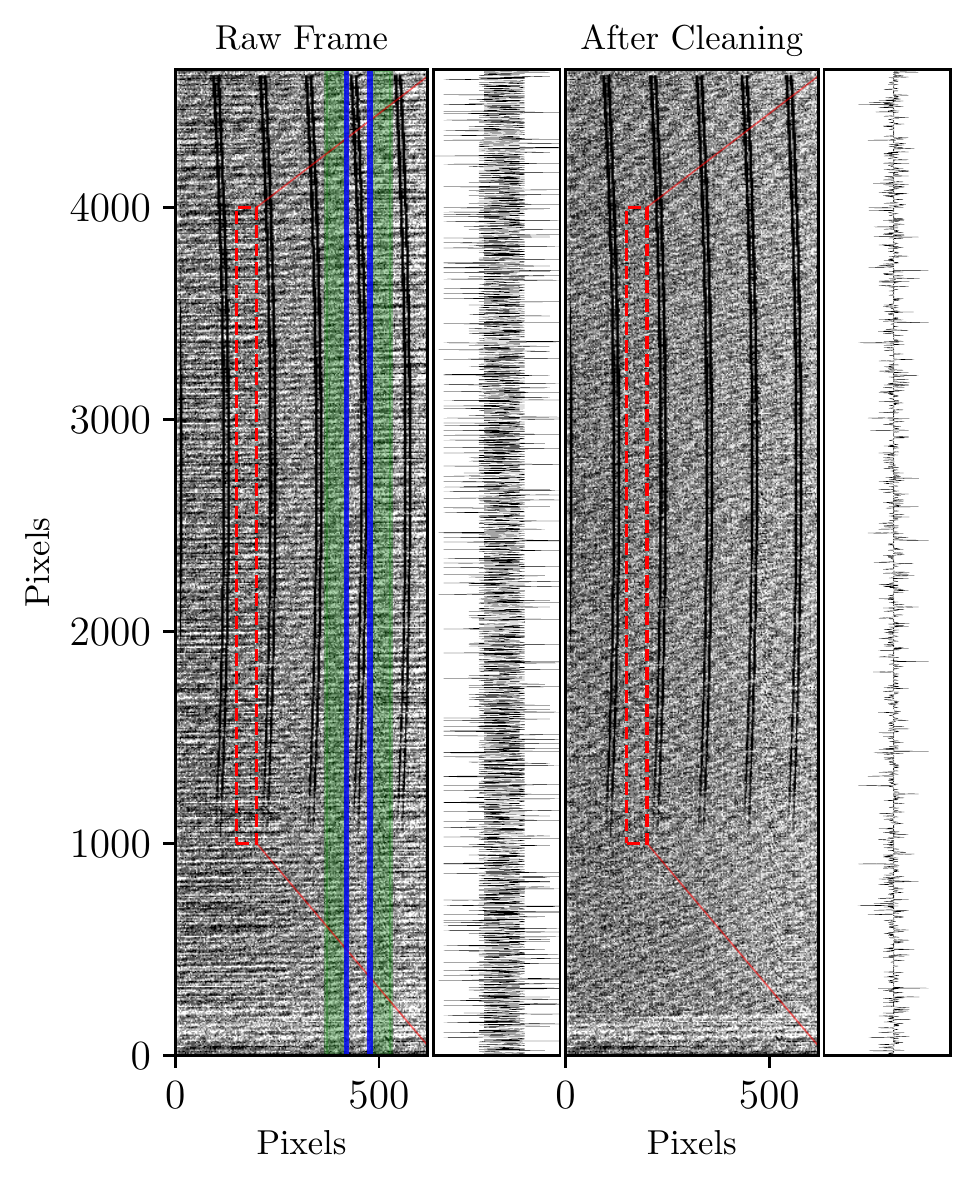}
    \caption{The results of the cleaning algorithm applied to a raw ESPRESSO frame. The left panel shows one of the readout sectors of a raw science frame and the right panel presents the same after the application of the algorithm. The additional boxes on the right sides show the median of the columns in the red dashed boxes, where the flux axis (x-axis) is set to the same scale in both plots; namely 1005 -- 1030\,ADU. Note that the first and last few rows are untouched by the algorithm, as to not modify any of the pre- and/or post-scan reads. This selection is much larger at the bottom where there are no traces present, whereas the reverse is true for the readout ports on top of the detector. The solid blue lines represent the location of slice columns detected by the algorithm, and the shaded green regions highlight the pixels from which the noise model is calculated.}
    \label{fig:Noise cleaning}
\end{figure}
In order to mitigate the impact of the noise generated by a combination of readout electronics and inhomogeneities in the coud\'e train, also noted by a number of previous studies \citep[e.g.][]{Allart2020,Tabernero2020,Borsa2020,Casasayas2021}, we wrote an algorithm to clean the raw frames before feeding them to the dedicated pipeline. This approach is preferred to modeling the noise after the reduction process, either in the stellar spectra or the differential residual spectra, as the noise pattern is linear in pixel space, and not wavelength. In other words, the noise pattern is observed to be more or less orthogonal to the CCD columns, where the echelle orders are curved. As the two ESPRESSO detectors are divided into 16 readout channels, where each has its own noise characteristics, we apply the cleaning procedure separately to each readout port\footnote{The exact details of the ESPRESSO readout electronics can be found in section 4.3 of the instrument manual (\url{https://www.eso.org/sci/facilities/paranal/instruments/espresso/ESPRESSO_User_Manual_P107.pdf}).}. In order to 
somewhat mitigate this noise, 
we go through an algorithm that performs the following steps:
\begin{itemize}
    \item detect columns that include pixels from any echelle slices;
        \begin{itemize}
            \item fit double-peaked Gaussian profiles for all lines of each port,
            \item for each slice-pair, identify the lowest column value on the left that is three FWHM away, as well as its mirrored value on the right (shown as vertical blue lines in left panel of Fig. \ref{fig:Noise cleaning}),
            \item build a pixel map of columns where echelle slices are present from the above values.
        \end{itemize}
    \item calculate the median of 50 pixel columns on both sides of the slices, while avoiding overlap with neighbouring slice columns (examples shown as green shaded regions in the left panel of Fig. \ref{fig:Noise cleaning}),
    \item subtract the mean of these median models to calculate a correlate-noise-only estimate for the columns that include echelle slices,
    \item subtract this model from the echelle slice columns, which leaves the underlying shot noise intact,
    \item apply this subtraction process to all other columns (where no echelle slices are recorded).
\end{itemize}
An example of one detector readout channel before and after the application of this cleaning algorithm is shown in Fig. \ref{fig:Noise cleaning}, where the efficacy of the algorithm in removing the dominant correlated noise is apparent in the flux boxes on the right side of each main panel. This approach is sensitive to both low and high frequency noise patterns noted by \citet{Allart2020}.

All ``cleaned'' frames were then reduced using the ESPRESSO data reduction workflow (pipeline version 2.2.1\footnote{\url{ftp://ftp.eso.org/pub/dfs/pipelines/instruments/espresso/espdr-pipeline-manual-2.2.1.pdf}}) provided by ESO and executed via the EsoReflex (version 2.11.0) environment. The reduction includes bias and dark subtraction, flat-field and bad pixel corrections, as well as cosmic ray removal. The traces are then extracted using an optimal extraction algorithm \citep{Horne1986}. The wavelength calibration is performed using arc frames taken with a Th-Ar lamp, as well as Fabry--P\'erot interferometer frames. 

\section{Analysis methods}
\label{sec:Methods}
\subsection{Telluric Correction}
\label{sec:Telluric}
For the purpose of modeling and correcting telluric transmission function, we use ESO's {\tt molecfit} routines \citep{Kausch2015,Smette2015}, version 1.5.9. It synthetically models telluric absorption lines using a line-by-line radiative transfer model. In order to estimate telluric line shapes and depths, the code requires the atmospheric pressure profile, which it calculates from the measured temperature and humidity profiles that it obtains from GDAS\footnote{Global Data Assimilation System}. This database provides profiles on a grid of 1$^\circ$ by 1$^\circ$, whose nearest data points to Cerro Paranal lie many kilometers away. {\tt molecfit} subsequently interpolates profiles among the four nodes around Paranal \citep{kimeswenger2015} and calculates the initial estimate of the pressure profile, using the hypsometric relation \citep[e.g.][]{Holton2013}. The telluric absorption model is then obtained through finding the best fitting pressure profile. As {\tt molecfit} expects one-dimensional spectra, one typically uses the S1D (linearised and stitched one-dimensional) spectra produced by the pipeline. However, we marginally improve upon this approach by creating our own S1D spectra, where the original sampling of the spectrograph is preserved\footnote{This is done by first combing the two slices of each order through resampling both slices onto the merged wavelength grid and taking the weighted average. The same procedure is then applied to the overlap regions of all neighbouring orders.}. This is done to enable the fitting algorithm to better model the resolution of the spectrograph. We do not use the sky-subtracted spectra as this additional step comes at the cost of adding noise to the final spectra and the observations were designed to avoid contamination from moon light. The very few sky emission lines are removed by our masking procedure during the data analysis (c.f. Section \ref{sec:processing}) and not considered in the modeling by {\tt molecfit}.

\begin{figure}
	\includegraphics[width=\columnwidth]{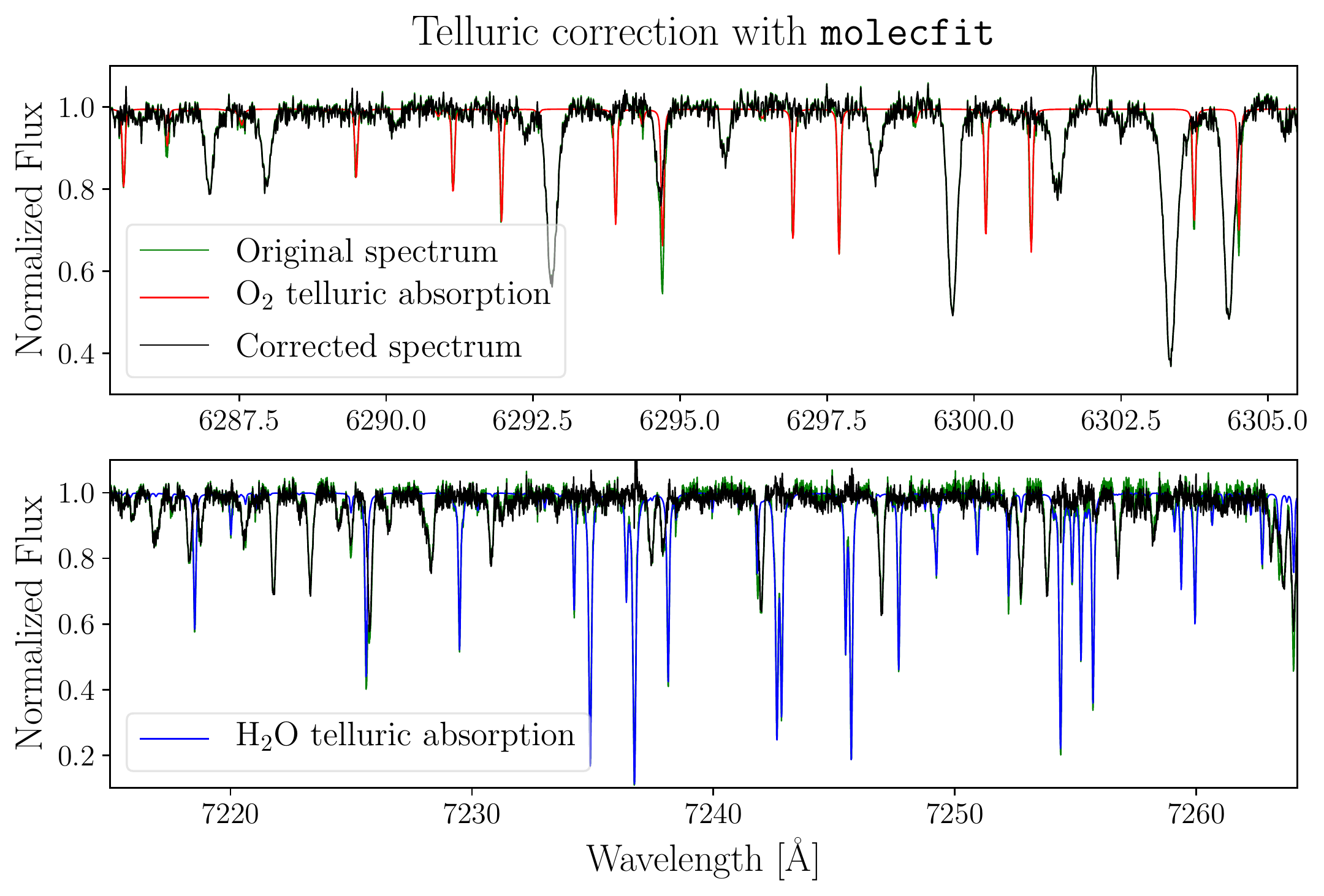}
    \caption{The process of correcting spectra for telluric absorption with {\tt molecfit}. \textit{Top} panel shows an O$_2$ region and the \textit{bottom} panel shows a section corrected for H$_2$O, with the wavelengths given in the observatory frame. After the telluric correction, the spectra are moved back to the barycentric frame to mitigate the radial velocity shift induced by Earth's barycentric motion.}
    \label{fig:Telluric}
\end{figure}

For those data sets where the data was available (namely DS4), we instead provided the pressure profile directly measured above Paranal, along the line of sight of the corresponding UT, by using the humidity and temperature profiles from the RPG\footnote{\url{https://radiometer-physics.de}} radiometer installed at Cerro Paranal that utilizes the LHATPRO\footnote{Low Humidity And Temperature PROfilers} model. The pressure profile is again derived using the hypsometric equation and calculated recursively at each height level. Using this method means that there is no need for fitting for the pressure profile, which vastly speeds up the code and marginally improves the precision of the final fitted transmission model. We fit for O$_2$ and H$_2$O absorptions only, as they are the main absorbers present in the ESPRESSO domain. We also fit for the resolution of the spectrograph via a variable kernel. As the reduction pipeline only provides wavelengths in the solar system barycentric rest frame, and {\tt molecfit} expects observatory frame wavelengths, we reverted back to this frame using the Barycentric Earth Radial Velocity (BERV) information given in the headers. Examples of telluric transmission models for both fitted species, as well as the corrected spectra are shown in Fig. \ref{fig:Telluric}.

We systematically created parameter files for each individual spectrum and ran the full set of {\tt molecfit} routines independently for each spectrum. Subsequently, telluric contamination is removed from the non-blaze-corrected S2D spectra (those that separately include flux information for individually extracted slices), that are used for the analysis in this work, by dividing them with the transmission model derived from the procedure described here. This approach has been shown to be more accurate in removing the telluric signature from high resolution spectra \citep{Langeveld2021}.

\subsection{Possible contamination from the 4LGSF}
\label{sec:LGSF}
As of April 2016, UT4 has been equipped with the 4 Laser Guide Star Facility (4LGSF), as part of the Adaptive Optics Facility (AOF), at the VLT. The vacuum wavelength of the laser emission line\footnote{Strictly speaking there are 3 laser emission lines; 18W at 5891.59210\,\AA, 2W at 5891.57137\,\AA~and 2W at 5891.61103\,\AA~\citep{Vogt2019}. However, the convolution with the instrumental profile (IP) and atmospheric broadening, means that the 4LGSF appears, approximately, as a single Gaussian to ESPRESSO.} is 5891.5912\,\AA, which lies close to the core of the Na\,{\scriptsize I\,D$_2$} line of the sodium doublet absorption, exact distance depending on the radial velocity of the star. From the sky projected distance between the field of view of ESPRESSO and the Rayleigh cone of the 4LGSF, one can expect contamination in the recorded spectra due to scattering and excitation in the atmosphere. Due to the proximity of our field of view to the Rayleigh cone of the 4LGSF, we note a possible contamination in DS1, and a possible additional enhanced sky glow in DS4, perhaps due to scattering from the UT4 dome. It must be noted that the angular separation between our pointing and the Rayleigh cone of the laser was larger than the zone of avoidance that is implemented in the automatic laser collision warning on Paranal. This suggests that, depending on the specific geometric constellation of target pointing and laser direction, indirect optical paths (e.g. multiple reflections) can cause additional contamination.

\begin{figure}
	\includegraphics[width=\columnwidth]{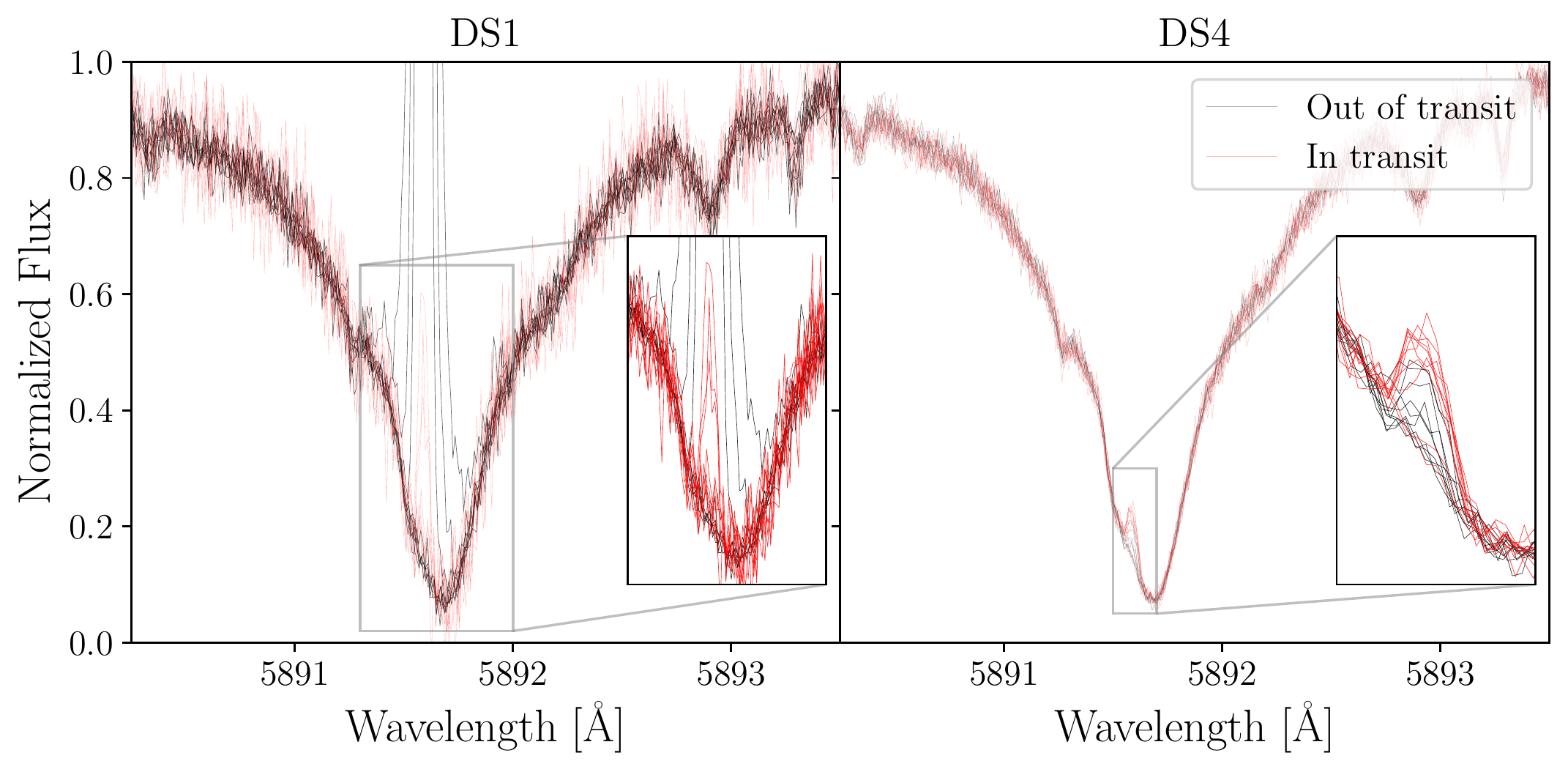}
    \caption{Contamination of the  Na {\scriptsize I D$_2$} line with the 4LGSF.  On the left the severe contamination outside of transit in DS1, as well as contamination of some in-transit spectra are evident.  On the right (in DS4), mild contamination of both in and out of transit spectra are shown.  In both cases, the level of contamination is far beyond any potential planetary transmission signal.}
    \label{fig:4LGSF}
\end{figure}

The contamination is at different levels for the two affected data sets, and impacts in- and out-of-transit spectra in an inhomogenous manner. Fig. \ref{fig:4LGSF} presents zooms into the {\scriptsize D$_2$} doublet region of the two affected data sets, where contamination is quite evident.

This possible contamination is particularly problematic for the calculation of individual line transmission spectroscopy of the sodium doublet. Its impact upon out of transit spectra can somewhat be mitigated by taking a median instead of a weighted mean, in the region of contamination. However, since the in-transit spectra are, initially, individually analyzed (see Sec. \ref{sec:TS}) and shifted to the stellar rest frame, the impact of the laser contamination upon residual spectra, and therefore the transmission spectrum is severe and very difficult to impossible to correct.  

Regardless, we made several attempts at correcting for this contamination by first scaling the sky spectrum, which also includes the laser line, and dividing it out. The scaling is necessary since the levels of contamination in the two fibers are significantly different. However, the uncertainties in this approach were orders of magnitude larger than any potential planetary transmission signal. We also experimented with modeling the individual laser emission lines, which again was not sufficiently precise. Therefore, it is concluded that the Na\,{\scriptsize I\,D$_2$} wing of the doublet is not available to us for any possible detection of atmospheric sodium, for the two affected data sets. The impact of this laser contamination has also been annotated in Fig. \ref{fig:TS-atomic}.

\subsection{Stellar \& orbital parameters}
\label{sec:Star model}
In order to determine stellar parameters, we took the weighted mean of all out-of-transit spectra in DS4, as this is the data set with the most precise spectra, and used that as the ``clean'' stellar spectrum. The {\tt zaspe} code \citep{Brahm2017} was then used to fit the spectrum, which coupled with results from GAIA DR2 \citep{Gaia2016,Gaia2018} and stellar isochrones from PARSEC \citep{Bressan2012}, allowed for calculation of the stellar parameters, given in Table \ref{tab:fit_results}. These sets of values are consistent with previously derived ones \citep{Hebb2010,Hellier2011,Torres2012,Tregloan-Reed2013}, with the exception of the age of the star which typically is not determined accurately from spectral synthesis alone, whereby results from asteroseismology would be much better trusted.

\begin{figure}
	\includegraphics[width=\columnwidth]{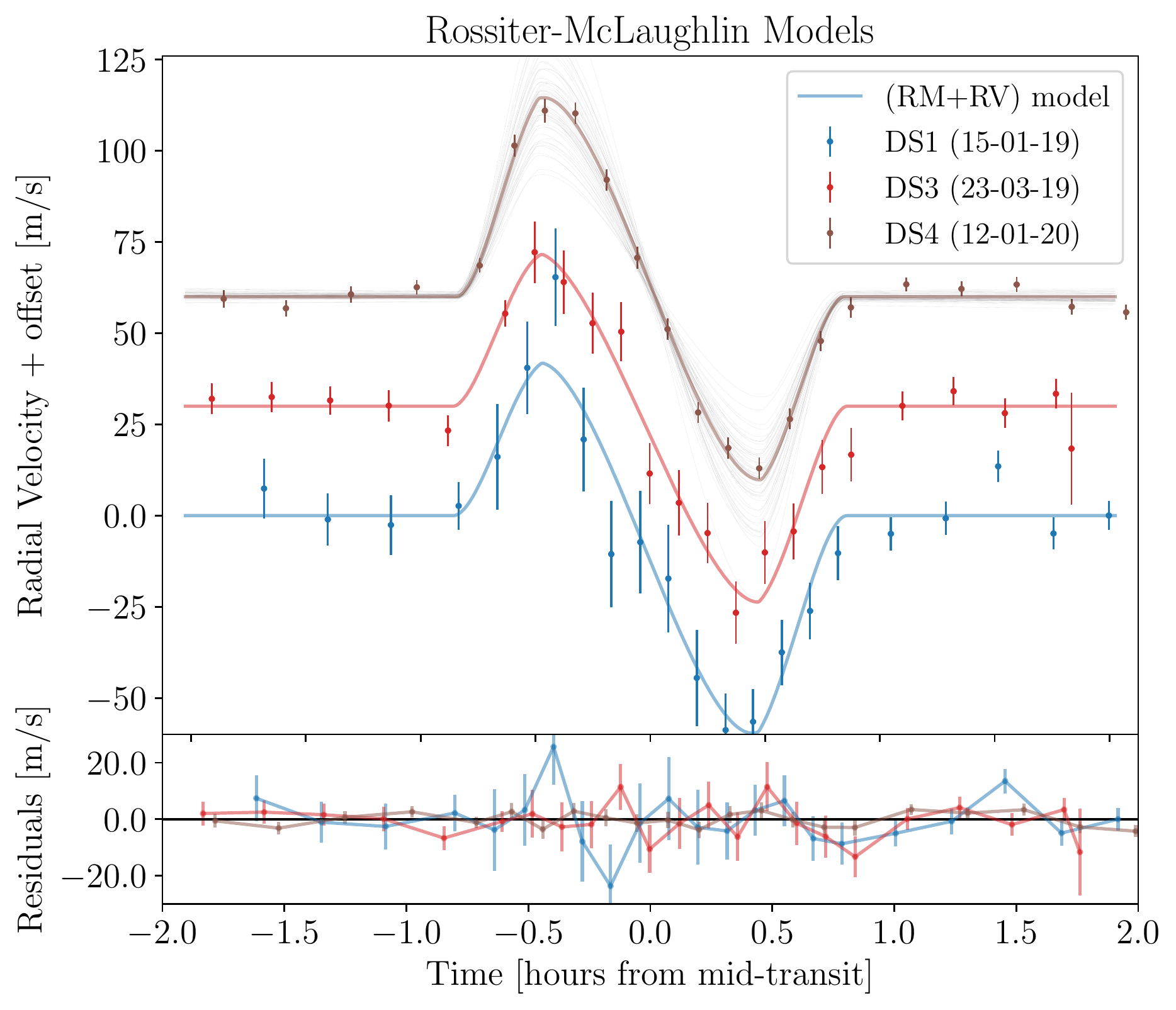}
    \caption{RV variations during three observed transits of WASP-19b (DS2 has been omitted due to the presence of high scatter). RV values (given in tables \ref{tab:RV1} to \ref{tab:RV4}) are determined by the reduction pipeline, through calculation of the Cross Correlation Function (CCF) with a G8 binary template (mask) matching the spectral type of the target. In all cases the underlying RV variations of the star (due to the presence of the planet) has been subtracted for better visualisation of the data and models. DS3 \& DS4 have been shifted vertically, by $+30$ and $+60$ m/s respectively, for clarity. The solid lines are best fit RM models, with the thin grey lines in DS4 presenting random draws from the posterior distributions of the RM model parameters. The lower panel presents the residuals of the best fit RV+RM models to the three data sets analysed.}
    \label{fig:RM}
\end{figure}

To determine various orbital and planetary parameters, we fitted the pipeline-measured radial velocity (RV) values of DS1, DS3 \& DS4 (calculated through cross-correlation with a G8 binary template and given in Appendix \ref{appendix:RV data}, tables \ref{tab:RV1}, \ref{tab:RV3} and \ref{tab:RV4}) with an RV model that includes the Rossiter-McLaughlin (RM) anomaly during transit \citep{Rossiter1924,McLaughlin1924}. The RV curve in DS2 (whose data are given in table \ref{tab:RV2}) is too noisy, due to poor observing conditions (see table \ref{tab:Observations}), and this data set is subsequently omitted from the remainder of this study. The composite model was created using formulations in {\tt PyAstronomy}'s \citep{Czesla2019} {\tt modelSuite} sub-package. Specifically, for the Keplerian RV variations, a circular approximation is assumed, and the \citet{Ohta2005} analytical description of the RM effect is employed\footnote{A caveat to note here is that this formulation has been shown \citep{Palle2020} to model best RM measurements derived from template matching approach \citep{Butler1996}, whereas the RM analytical definition of \citet{Boue2013} is best suited to fit values derived from the CCF technique employed by the DRS. However, a comparison between parameter values derived from this modeling and those previously reported \citep{Hellier2011,Tregloan-Reed2013} shows that this caveat is of minimal importance.}. In all three cases, we modeled the RV curve using both this model and a second model where a linear trend in time is also added to account for the impact of stellar activity upon the slope of the underlying RV curve. The theoretical impact of stellar activity upon the measurement of the RM effect is extensively discussed in \citet{Boldt2020}. 
The fitted and fixed parameters of our model are given in table \ref{tab:fit_results}, with the exception of the systemic velocity, $\nu_{\textrm{sys}}$, whose value was allowed to vary within a 40\,m/s box around the known value for the star. The individual results for this parameter are omitted from table \ref{tab:fit_results}, as its posteriors do not show clear distributions, and subsequently it is treated as a nuisance parameter. We initially attempted a joint fit across the 3 data sets;  however, convergence could not be achieved for most orbital parameters. Subsequently, we fitted each curve individually with the best fit RM models shown in Fig. \ref{fig:RM}, where the underlying RV model has been subtracted for better clarity. We report in table \ref{tab:fit_results} the fitted RM parameter values from the analysis of DS4. This was done by running multiple Markov Chain Monte Carlo (MCMC) simulations of 200k iterations, with the first 5000 steps taken as burn-in. We checked the posterior distributions of the fitted parameters (presented in Fig. \ref{fig:RM4 posteriors}) for convergence and quote the final results from the mean and 16/84 percentiles.  \textit{It must be noted that there was no statistical evidence for the inclusion of the additional slope due to activity, in any of the data sets analysed}.

We note a small offset 
in the determined systemic velocity between the two 2019 data sets (DS1 \& DS3) and the one taken later in 2020 (DS4), which is attributed to a combination of the exchange of the fiber-link in July 2019, 
the different readouts of the detector employed\footnote{This has been shown to lead to a $\sim$2.2\,m/s systematic offset in the RV measurement (\url{eso.org/sci/facilities/paranal/instruments/espresso/ESPRESSO_User_Manual_P102.pdf}).} and the uncertainty in determination of its value from the modeling process above. Analysis of activity levels of the star during the observations, through the calculation of both the S-index and $\log R^{\prime}_{HK}$ \citep{Buccino2008}, does not show any significant variation, shown in Fig. \ref{fig:Activity}.

Additionally, the determination of stellar rotation velocity $\nu \sin{i}$ from the stellar template analysis and the RM fitting, yield two significantly different values. This difference is attributed to how each approach is affected by macro-turbulence in the stellar atmosphere, where in the spectral template approach, the measurement of line-broadening includes contributions from velocity fields in the stellar photosphere \citep{Doyle2014}.  

\begin{table}
 \caption{Summary of various stellar, orbital and planetary parameters derived from the analysis of the stellar spectra and the modeling of the Rossiter-McLaughlin effect.}
 \label{tab:fit_results}
 \begin{tabular}{lll}
  \hline
  Parameter &  Prior & Value\\
  \hline \hline
  Orbital period, P [days] & \textit{Fixed} & 0.788839$^a$\\[2pt]
  Orbital eccentricity, \textit{e} & \textit{Fixed} & 0.0$^a$ \\[2pt]
  \hline
  \multicolumn{3}{c}{\textit{Stellar spectrum analysis$^b$}}\\[2pt]
  Age [Gyr] & -- & 8.9 $^{+2.5}_{-2.8}$ \\ [2pt]
  Mass, M$_\star$ [M$_\odot$] & -- & 0.92 $^{+0.04}_{-0.03}$ \\ [2pt]
  Radius, R$_\star$ [R$_\odot$] & -- & 0.98 $\pm$ 0.01 \\ [2pt]
  Effective temperature, T [K] & -- & 5497 $\pm$ 70 \\ [2pt]
  Luminosity, L$_\star$ [L$_\odot$] & -- & 0.809 $^{+0.044}_{-0.035}$ \\ [2pt]
  Metallicity, [Fe/H] [dex] & -- & 0.12 $\pm$ 0.04\\ [2pt]
  Surface gravity, log (g) [cgs] & -- & 4.421 $^{+0.024}_{-0.020}$ \\ [2pt]
  Rotation velocity, $\nu \sin{i}$ [km/s] & -- & 4.59 $\pm$ 0.30\\[2pt]
  V-band extinction, $A_\text{V}$ [mag] & -- & 0.125 $^{+0.079}_{-0.067}$ \\ [2pt]
  \hline
  \multicolumn{3}{c}{\textit{........................................    RM analysis -- DS1,3,4    ........................................}}\\[2pt]
  \textbf{1} T$_0$ [BJD+2458498] & $\mathcal{U}(0.63,0.69)$ & $0.6626 \pm 0.0012$\\[2pt]
  \textbf{3} T$_0$ [BJD+2458565] & $\mathcal{U}(0.68,0.74)$ & $0.7106 \pm 0.0008$\\[2pt]
  \textbf{4} T$_0$ [BJD+2458860] & $\mathcal{U}(0.71,0.77)$ & $0.7375 \pm 0.0003$\\[2pt]
  \multicolumn{3}{c}{\textit{.......................................    RM analysis -- DS4   ........................................}}\\[2pt]
  Rotation velocity, $\nu \sin{i}$ [km/s] & $\mathcal{U}(2.0,10.0)$ & 6.40 $^{+0.9}_{-1.0}$\\[2pt]
  Linear limb-darkening, \textit{u} & $\mathcal{U}(0.0,1.0)$ & $0.91 \pm 0.05$\\[2pt]
  Stellar axis inclination, I$_\star$\,[$^\circ$] & $\mathcal{U}(50,90)$ & $69 \pm 14^c$\\[10pt]

  Semi-major axis, a [$R_\star$] & \textit{Fixed} & 3.5875$^d$\\ [2pt]
  RV semi-amplitude, $K_\star$ [m/s] & \textit{Fixed} & 257.0$^a$\\ [2pt]
  Orbital inclination, \textit{i}, [$^\circ$] & \textit{Fixed} & 79.52$^d$\\ [2pt]
  Projected spin-orbit, $\lambda$ [$^\circ$] & $\mathcal{U}(-10,15)$ & $-1.9 \pm 1.1$\\ [10pt]
  
  Planet radius, R$_\text{p}$ [$R_\star$] & $\mathcal{U}(0.130,0.160)$ & $0.1449 ^{+0.0098}_{-0.0096}$\\ [2pt]
  \hline
 \end{tabular}
  \vspace{1ex}
     {\raggedright $^{(a)}$ \citet{Hellier2011}. \\
     $^{(b)}$ All reported values are from the spectral analysis with {\tt zaspe}.\\
     $^{(c)}$ Posterior con constrained, c.f. Fig. \ref{fig:RM4 posteriors}.\\
     $^{(d)}$ \citet{Sedaghati2017}.\par}
\end{table}

\subsection{Post processing and cross correlation}
\label{sec:processing}
From the pipeline products, we choose to strictly work with the non-blaze-corrected S2D spectra. This choice is motivated by two reasons. S2D is chosen over S1D as it contains the native sampling of the spectrograph and consequently involves no reinterpretation of flux, done via resampling. Furthermore, this choice allows for calculation of the Cross Correlation Function (CCF) as a function of echelle orders (as is done by the pipeline in the calculation of the RV), which allows greater flexibility in the interpretation of any atmospheric detections. Secondly, we choose those spectra that still contain the blaze function as this negates the need for expensive weighting steps when either masking points or calculating the CCF. This approach ensures the retention of absolute flux, which implicitly weighs each pixel according to its variance when performing either of the two aforementioned operations. This is similar to the approach that was adopted by \citet{Hoeijmakers2020} in the analysis of HARPS data of WASP-121.

In order to enable vectorization of various processing algorithms and therefore greatly speeding up calculations, the flux and wavelength information for each data set are read into a matrix for each given order individually, where this M\,$\times$\,N matrix has columns (M) equal to the number of pixels in each order and rows (N) equal to the number of observations in that data set. Of the 170 slices of the echellogram, we ignore the bluest 44 slices, where the measured flux is completely dominated by shot noise, meaning a starting wavelength of $\sim$\,4370\,\AA~for the spectra. We then perform these following steps in order to prepare the spectra for further analysis:

\begin{enumerate}[label={(\arabic*)}]
    \item \textbf{Telluric correction:} The derived telluric transmission models are arranged into the same M\,$\times$\,N matrix and resampled onto the wavelength matrix, for each slice. The corresponding elements of the two matrices are then divided to remove the telluric absorption. Additionally, any of the columns in the telluric matrix, where more than 20\% of the values are $>$50\% deep are flagged, and values in their corresponding columns in the flux matrix are replaced by NaN's (not a number). This ensures that the regions of deep telluric lines (mostly in the O$_2$ bands) do not contribute unnecessary noise to the CCF.
    \item \textbf{Shifting spectra:} We put all the spectra in a common rest frame in which the star has no reflex motion. This is done by calculating the stellar radial velocity relative to mid-transit and applying a doppler shift to the rows of the wavelength matrix accordingly.
    \item \textbf{Outlier removal:} We identify sky emission lines or imperfectly corrected cosmic rays by flagging those pixels that are 4$\sigma$ above the continuum, where the flux standard deviation is calculated individually in each slice and only in the continuum. The continuum itself is modeled through the product of the blaze function (produced by the pipeline for each observation individually) and a linear polynomial. We then replace a flux column by NaN's if more than 20\% of that column has been flagged. Otherwise, the outlier is replaced by the value of the calculated continuum at that wavelength.
    \item \textbf{Noise suppression:} We further mask those columns where the S/N ratio is below 0.5 and replace them with NaN's. This step essentially masks out the deepest stellar cores, as well as the very noisy edges of slices, where there is very little astrophysical signal retained. This step is only performed for those spectra that are used in the calculation of the CCF, and not for narrow-band spectroscopy.
\end{enumerate}

The above steps mask out a total of 3.39\% of all flux points produced by the pipeline. The flux and wavelength matrices are then saved for each slice as extensions of a fits file and stored for further analysis.

As the principal approach, we use the cross correlation technique \citep{Snellen2010} to search for absorption signatures emanating from the exoplanetary atmosphere, for neutral atomic and molecular species. The cross correlation of a spectrum of fluxes $f_i$ with a template $T_i$ is simply defined as the dot product of the two arrays, $f \bullet T$. We calculate the matrix of CCF's, $C$, for the flux arrays described above $f_i(t)$ with a template $T$, Doppler-shifted to a velocity $\nu$ and resampled onto the wavelength grid $\lambda(t)$ as:
\begin{equation}
    \label{eq:CCF}
    C(\nu,t) = \sum_S \sum_{i=0}^{N} \left[ f_i \Big( t,\lambda(t) \Big) * T_i \Big( \nu, \lambda(t) \Big) \right ]_s
\end{equation}
\noindent where $S$ represents all the slices for which the cross correlation is performed, whose arrays are to be counted towards the final sum, $N$ is the total number of flux values in a given slice and the mathematical operation inside the sum denotes element-wise multiplication of the two matrices. The sum produces a vector of cross correlation values, the size of which is equal to the number of exposures. Scanning this CCF as a function of velocity $\nu$ results in the CCF matrix. As the wavelength arrays for the spectra at different times are not identical due to the slight shifts applied because of the reflex motion of the centre of mass of the system, we resample the template onto each row separately, and thereby avoid resampling of the spectra themselves. The vectorised nature of such operations in {\tt python} for matrices means that this comes at no extra cost in calculation time.

\subsection{Model spectra and cross correlation templates}
\label{sec:CCF templates}
We calculated model transmission spectra for the atmosphere of WASP-19b using the {\tt petitRADTRANS} package \citep{Molliere2019}, which is a radiative transfer code written for the purpose of modeling exoplanetary atmospheres. Using the high-resolution, line-by-line mode, we calculated spectra for a Hydrogen/Helium dominated atmosphere with everything else included as trace species. We used the planetary radius (1.3836\,$R_{\mathrm{\textit{jup}}}$), surface gravity ($\log (g) = 3.17$\,cgs) and atmospheric mean molecular weight (2.0159\,\textit{amu}) values determined by \citet{Sedaghati2017} and employed the provided utility function from \citet{Guillot2010} to analytically calculate the P-T profile of the atmosphere. Additionally, we included continuum opacity of collision induced absorption from H$_2$-H$_2$ and H$_2$-He, as well as Rayleigh scattering from molecular hydrogen. The reference pressure, the pressure above which the atmosphere is opaque at all wavelengths, is also set to the value determined from the low resolution FORS2 transmission spectrum of 0.93\,bar. The equilibrium temperature is also set to the value of 2350\,K derived from the low-resolution atmospheric retrieval. We assumed a uniform abundance distribution of the trace species with pressure and therefore no chemistry or dynamics are taken into account. Models are calculated for abundances ranging from sub- to super-solar, which are presented in section \ref{section:CCF analysis}.

We systematically searched for the presence of Fe\,{\scriptsize I}, H$_2$O and TiO, by calculating transmission spectra for those individual species, for which the line lists are obtained from the HITEMP \citep[H$_2$O;][]{Rothman2010}, ExoMol \citep[TiO;][]{McKemmish2019}, as well as Kurucz \citep[Fe;][]{Kurucz2017} databases. Once calculated, the transmission models are convolved with a Gaussian kernel in order to match them to the resolution of the spectrograph. The wavelength dependent Instrumental Profile (IP) is calculated by fitting Gaussian functions to all strong emission lines of Thorium Argon frames taken for the purpose of wavelength calibration. FWHM of singular lines are used to determine the dependence of the IP on wavelength, which is then modelled as a linear polynomial. The convolution of spectra is then performed with a Gaussian kernel of a variable width determined by this function.

These transmission spectra of individual species are used for both model injection and cross correlation analysis. In both cases the continuum is traced with a high bandpass filter and modeled using a high order polynomial, where for the CCF template this model is subtracted from the spectrum. This ensures that the data points where there is no absorption from the trace species do not contribute to the calculation of the final CCF. Additionally, any values below 0.5\% of the largest peak are also set to zero to minimize noise in the CCF and the template is finally normalised by its own sum, namely $\sum T = 1$. This normalisation step is performed separately for every RV step at which the template is calculated.

The calculated transmission spectra are also used for the purpose of model injection and retrieval in order to determine the abundance limit at which the data allow for detection of a given species \citep[e.g.][]{Birkby2013,Hoeijmakers2015}. This is done by normalising the IP-convolved spectra to unity, through dividing by the continuum model and applying two further broadening kernels. The first is to account for the change in radial velocity of the planet from the start to the end of each exposure, following the approach of \citet{Brogi2016}. This is done through convolution with a box kernel whose width matches this change in radial velocity, performed individually for injections at each exposure. The second is to account for the rotation of the tidally locked planet ($\nu_{\textrm{eq}} = 9.37$\,km/s) and the inclination angle of the orbit, where for rotation broadening it is given by $2\sqrt{\ln{2}}\lambda\langle\nu\rangle\sin{i}$, with $\langle\nu\rangle\, (= 2\nu_{\textrm{eq}}/\pi)$, the average rotation velocity of the atmosphere from equator to pole. This convolution is performed with a Gaussian kernel of width matching this average rotation velocity. However, this second broadening step is under the assumption that the upper atmosphere rotates synchronously with the planet, which almost certainly is not the case \citep[e.g.][]{Showman2011a,Showman2011b,Showman2013,Heng2015}. These models, calculated at a range of abundances, are finally injected into the spectra via multiplication, at a velocity anti-directional to the orbit of the planet and a $\nu_{sys}$ offset from the system. This is done in order to minimize any overlap between the injected signal and any possible planetary signal in the velocity-velocity maps, described in Section \ref{section:CCF analysis}.

\subsection{CLV+RM modeling}
\label{sec:CLV+RM}
\begin{figure*}
	\includegraphics[width=\textwidth]{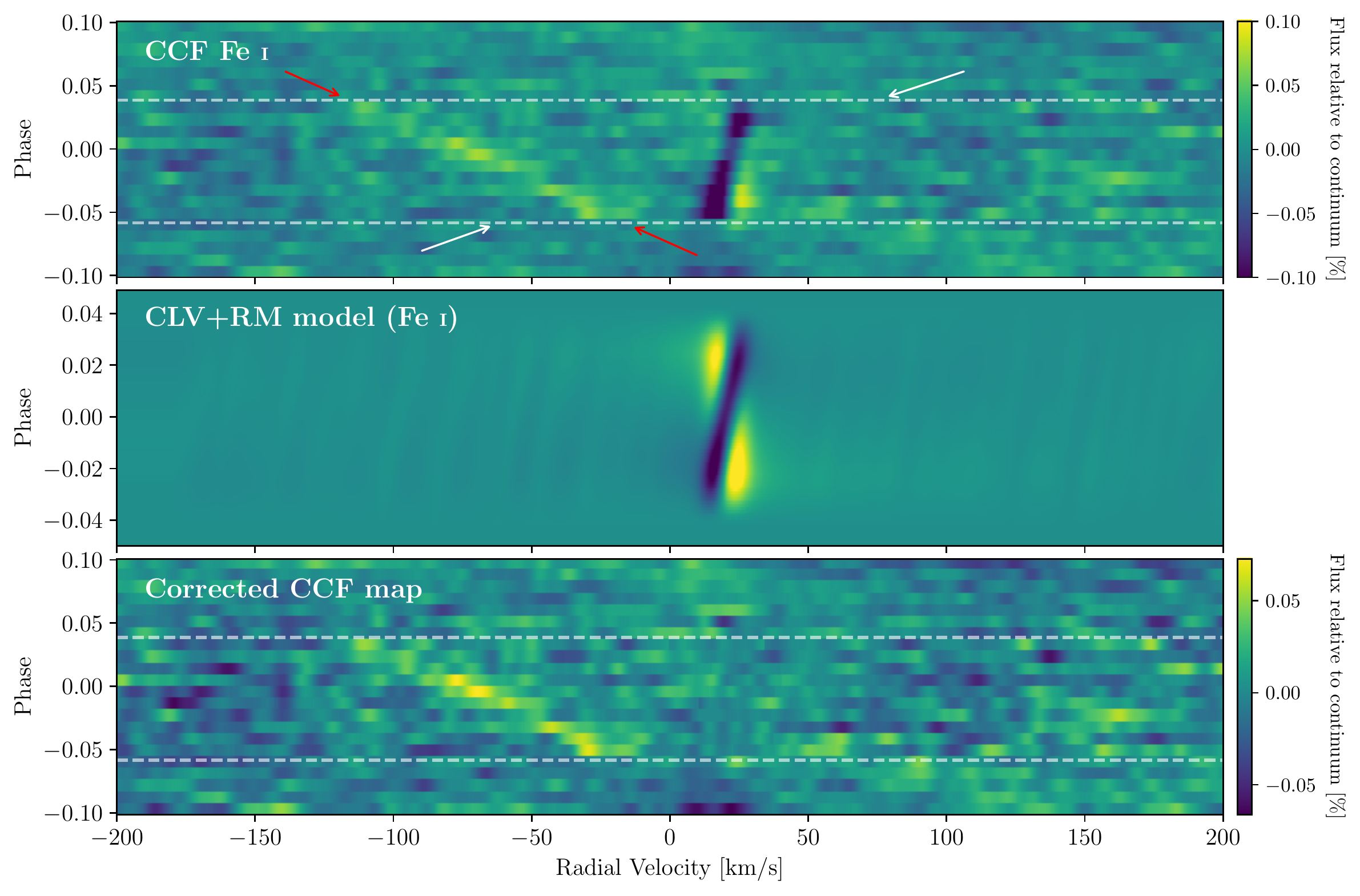}
    \caption{The process of modeling and removing the CLV+RM effect from the CCF maps. \textit{Top} panel shows the map for the cross correlation of the observed data with a template of Fe {\scriptsize I}. The horizontal dashed lines are the transit contact points. The white arrows indicate the trace of the planet for its calculated 228.24 km/s orbital velocity, and the red arrows show the velocity at which a signal is injected into the data. \textit{Middle} panel shows the modeled, combined CLV and RM effect, as described in Section \ref{sec:CLV+RM}. \textit{Bottom} panel shows the residuals when the scaled CLV+RM model is subtracted from the data.}
    \label{fig:CLV+RM}
\end{figure*}
Observations of exoplanetary transits with precision, high resolution spectroscopy involve a multitude of imprints upon different aspects of recorded spectra. In addition to absorption imprints from various atomic and molecular species in the atmosphere, there exists substantial deformation of stellar absorption lines. This deformity originates from the specific region of the visible stellar disk that is occulted by the transiting planet. As the projected disk of the planet moves from the limb to the centre of the star, there is a differential effect due to the limb-darkening of the stellar disk.  This phenomenon is known as Centre to Limb Variations (CLV) which manifests itself as an anomaly in both the calculated cross correlation map and the narrow-band transmission spectrum, that has to be accounted for by a theoretical modeling process. Additionally, the RM effect also has a similar impact on both analyses, which again has to be modeled and accounted for. For a more detailed discussion and analysis of both these effects the reader is encouraged to refer to \citet{Czesla2015}, \citet{Louden2015}, \citet{Yan2017}, \citet{Borsa2018} and \citet{Casasayas2020}.

For the modeling of both these effects we follow the steps taken by \citet{Casasayas2020}, a brief summary of which we present here. Initially we model a template for the host star using line lists from the upgraded {\tt VALD} database \citep{Ryabchikova2015,Piskunov1995}, {\tt ATLAS9} models \citep{Castelli2003}, calculated with the {\tt Spectroscopy Made Easy} tool \citep[SME;][]{Valenti1996}, while local thermodynamic equilibrium (LTE) is adopted. In order to calculate variations of the stellar spectrum during transit, i.e. the impact of the occulting disk of the planet, we divide the stellar surface into a grid of $0.01R_\star\times0.01R_\star$ resolution, where each cell has its own spectrum individually defined as a function of its limb angle $\mu$ and projected rotation velocity. 

For any given instance during transit, the stellar spectrum is the integral of flux from all non-occulted cells, where the planet is assumed to be an opaque disk of constant radius at all wavelength, i.e. no atmosphere is assumed. To calculate the impact of these two effects in narrow-band transmission spectroscopy (c.f. Fig. \ref{fig:TS-atomic}), the in-transit spectra are normalized to the out-of-transit modeled spectrum, that does not include neither CLV nor RM effects, shifted to the planetary rest-frame and subtracted out of the final transmission spectrum, described in section \ref{sec:TS}. This combined modeled effect is shown as a solid green line in some panels of Fig. \ref{fig:TS-atomic}, where the impact is found to be minimal, relative to any hypothetical absorption features.

To obtain the exact form of this effect in the cross correlation map for individual species, we cross correlate those synthetic models (described above that include only the effect and nothing else) with the atomic or molecular templates. We then bin these calculated CCF's to match the phases of individual exposures and subtract the out of transit CCF. This gives the expected CLV+RM effect in the cross correlation map, an example of which is shown in the middle panel of Fig. \ref{fig:CLV+RM}, calculated for Fe {\scriptsize I} template. To then remove this effect, we scale the amplitude of the model to match that of the observations and subtract it from the calculated CCF map of the observations (c.f. bottom panel of Fig. \ref{fig:CLV+RM}). The scale factor is estimated through a $\chi^2$ minimization approach.

\subsection{Narrow-band transmission spectroscopy}
\label{sec:TS}
In addition to the phase-resolved study of an exoplanetary atmosphere through the cross correlation technique, one can also probe the presence of a species through differential transmission spectroscopy of strong, individual absorption lines \citep[e.g.][]{Redfield2008,Wyttenbach2015,Wyttenbach2017,Casasayas2017,Casasayas2018,Nortmann2018,Zak2019,Chen2020}. In this approach, the stellar signature is removed from in-transit spectra. To obtain a stellar template, one that is devoid of any planetary absorption, we take a weighted mean of the out of transit spectra. The transit contact points are determined through the analysis of the RM effect for each transit individually. Due to the presence of the planet, these spectra have an intrinsic RV shift, which is corrected for by mapping all spectra onto the stellar rest frame. 

To do this, the 
RV is read directly from the pipeline for the out of transit spectra, and calculated for the in-transit spectra 
using the RV component of our composite RM+RV model described in section \ref{sec:Star model} (namely the {\tt KeplerRVModel} formulation from {\tt PyAstronomy}'s {\tt modelSuite}). This is to determine the true value of the 
stellar reflex velocity in-transit, whereby the pipeline calculated values are affected by the RM anomaly. To finally place all spectra in the stellar rest frame, they are Doppler corrected for the RV of the star relative to its systemic velocity, determined from the RM modeling. 

\begin{figure*}
	\includegraphics[width=\textwidth]{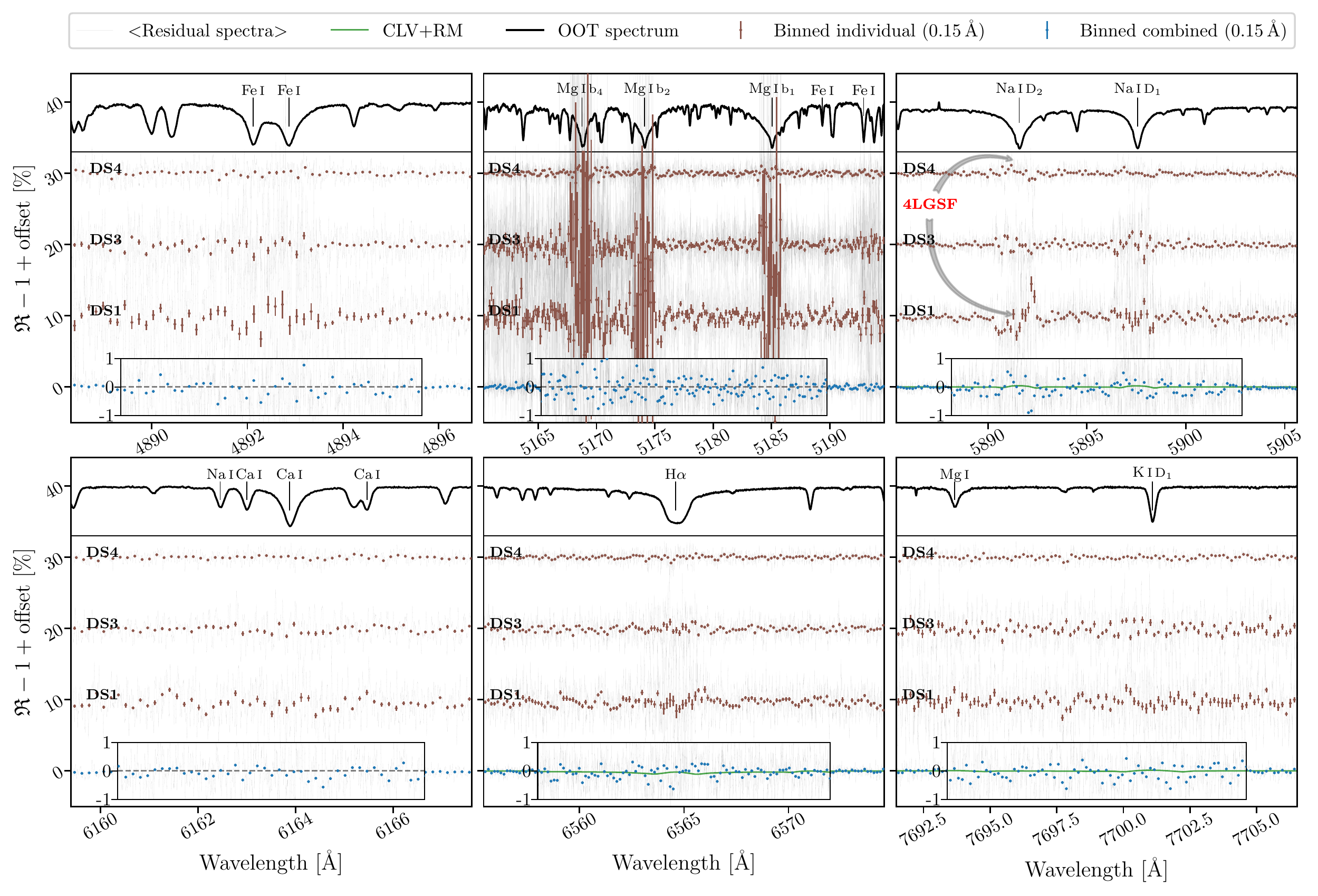}
    \caption{Transmission spectrum of WASP-19b at locations of various strong transition lines.  In each panel, the solid black spectrum at the top is the combined out of transit stellar spectrum, used for the removal of the stellar signature, where the thin horizontal line shows the zero flux level (the continuum and zero levels are arbitrarily set at 40\% and 33\%, respectively). The location of the line/s being probed, as well as some other lines, have been annotated at the top of each panel. All transmission spectra from the individual and the combined data sets are shown in thin gray, where the binned spectra are represented in brown and blue, for individual and combined data sets, respectively, all binned within 0.15\,\AA~bins. Additionally each of the individual transmission spectra have been annotated on the left hand side. The inset panel in each figure represents a zoom along the transmission axis, for a clearer inspection of the combined transmission spectrum. The modeled CLV+RM effect is shown as a green line for the combined transmission spectrum, only in three cases, where its minimal impact upon the final transmission spectrum is evident. All individual transmission spectra are shifted upwards recursively by 10\% for better visualization.}
    \label{fig:TS-atomic}
\end{figure*}

To create the out-of-transit template, we resample all spectra onto a common wavelength grid encompassing all individual wavelength samples, ensuring conservation of flux at every point. We then calculate the weighted mean of the out-of-transit spectra, where the weights of individual flux values are taken as the inverse of standard error squared. Sections of this out-of-transit stellar template, without the blaze function, are shown in the top panels of Fig. \ref{fig:TS-atomic}. This procedure is performed separately for individual data sets.

The transmission spectrum is subsequently created by removing the stellar signature, using the stellar template created above, from the in-transit ones, and then shifting the individual residuals to the planetary rest-frame. This is to correct for the radial velocity of the planet during transit, in order to shift any potential atmospheric absorption to a common frame. The residual spectra, $\textbf{r}(\lambda,t)$, are calculated by dividing all spectra (in and out of transit) by the weighted mean template spectrum. This step implicitly removes the blaze function of the spectrograph. 

The epoch-dependent planetary radial velocity is calculated via the relation $\nu_p = K_p \sin{2\pi \phi}$, where the phase $\phi$ is given as $(T_{\text{mid-exp}}-T_0)/P$ and the planetary RV semi-amplitude, $K_p = 228.24 \pm 7.45$ km/s, from conservation of momentum ($K_p = K_\star M_\star / M_p$). The stellar RV semi-amplitude ($K_\star$) is 
taken from \citet{Hellier2011}, stellar mass ($M_\star$) from spectral synthesis and the planetary mass ($M_p$) from the observed period and the Kepler's laws.

The final transmission spectrum $\mathfrak{R}$ is then calculated from the weighted sum of these shifted residual spectra:
\begin{equation}
    \mathfrak{R} = \sum_t \textbf{r}(\lambda,t) \cdot \Tilde{w}_t \; \; \; \; with \; \; \; \; \Tilde{w}_t = w_t / \Sigma_i (w_i)
\end{equation}
where $\textbf{r}(\lambda,t)$ are the residual spectra in the stellar rest-frame, and $w_t$ and $\Tilde{w}_t$ are the individual weights and normalized weights, respectively.  This approach is similar to the one taken by \citet{Wyttenbach2017} in producing the transmission spectrum from high resolution spectra, sections of which are shown in Fig. \ref{fig:TS-atomic}.

\section{Results}
\label{sec:Results}
\subsection{Transmission spectra}
We perform narrow-band transmission spectroscopy of WASP-19b for those species with strong singular transition lines, by scanning the order-by-order transmission spectrum, $\mathfrak{R}$, for significant absorption features.  This is done both for individual data sets, as well as their weight-combined averages. The final transmission spectra are further corrected for the combined CLV+RM effect described in Section \ref{sec:CLV+RM}, where the amplitude of the effect is determined to be not significant relative to the noise floor in the deferentially combined spectra. Some of such models are shown in panels of Fig. \ref{fig:TS-atomic} as solid green lines.

Searching through both individual and combined transmission spectra we do not detect any significant absorption present in any sets of observations. Any single data set detection is attributed to either stellar line variability (be it in the core or continuum) or residual systematic noise associated with the detectors, not fully compensated for by the procedure in section \ref{sec:electronic noise}. Although we search the entire spectral range, zooms into specific regions of interest are shown in Fig. \ref{fig:TS-atomic}. In each panel both the individual transmission spectra, as well as the combined spectrum are presented.

Subsequently probing a number of different absorption lines, we place upper limits on the contrast of any possible absorption from the planetary atmosphere. Those limits are presented in Table \ref{tab:atomic upper limits}, where upper limits on both the contrasts and core depths of any possible planetary absorption lines are derived from posterior Empirical Monte Carlo \citep[EMC;][]{Redfield2008} distributions (c.f. section \ref{sec:EMC}), for 1\,\AA~and 0.30\,\AA~bins centered on the core of a given line.

\begin{table}
 \caption{Summary of 1$\sigma$ upper limits placed on various core contrasts in the transmission spectrum of WASP-19b.}
 \label{tab:atomic upper limits}
 \centering
 \begin{tabular}{lccc}
  \hline
  Line &  Location & Contrast               & Core depth\\
       &  [\AA]    & ($\pm 0.5$\,\AA) [\%]  & ($\pm 0.15$\,\AA) [\%] \\
  \hline \hline
  H\textbeta & 4862.708 & $\leq 1.46\pm0.13$ & 0.71\\[2pt]
  Fe\,{\scriptsize I} & \textit{various} & $\leq 1.27\pm0.21$ & $\sim$\,0.21\\[2pt]
  Na\,{\scriptsize D$_2$} & 5891.583  &   \multicolumn{2}{c}{\textit{...................    4LGSF    ...................}}\\[2pt]
  Na\,{\scriptsize D$_1$} & 5897.558  & $\leq 0.99\pm0.08$ & 0.12\\[2pt]
  Na\,{\scriptsize I} & 6162.452  & $\leq 0.67\pm0.11$ & 0.09\\[2pt]
  Ca\,{\scriptsize I} & 6564.603  & $\leq 0.67\pm0.14$ & 0.07\\[2pt]
  H\textalpha & 6564.603 & $\leq 0.89\pm0.08$ & 0.19\\[2pt]
  K\,{\scriptsize D$_2$} & 7667.009  & \multicolumn{2}{c}{\textit{Blended with an O$_2$ telluric A-band.}}\\[2pt]
  K\,{\scriptsize D$_1$} & 7701.084  & $\leq 0.59\pm0.06$ & 0.07\\[2pt]
  \hline
 \end{tabular}
\end{table}

\subsubsection{Stellar activity indicator diagnostics}
\citet{Sasso2017} suggest the Mg\,{\scriptsize I\,b} Fraunhofer triplet, as well as the singular line at 4572\,\AA, as diagnostics for stellar chromospheric activity, where the singular line is determined to be more sensitive to changes in the atmospheric structure, as compared to the triplet. We analysed the transmission spectrum in both of these regions and find no systematic deviations relative to the noise floor, in either region for making any definitive conclusions about the activity levels. These lines have previously been used by other studies as control in high resolution transmission spectroscopy \citep{Wyttenbach2017,Zak2019}. Furthermore, we look at a Calcium triplet at $\sim$\,6164\,\AA~that has been suggested as indicative of chromospheric activity \citep{Houdebine2010} and find no variability in its transmission spectrum. This is shown in the left panel of the bottom row in Fig. \ref{fig:TS-atomic}.

Given both of these indicators, together with the fact that both the H\textalpha~line and Na\,{\scriptsize D} doublet could also be indicative of chromospheric activity \citep{Pasquini1991,Andretta1997}, given a substantial spot coverage of the stellar surface \citep{Cauley2018}, we can tentatively rule out stellar chromospheric activity as the cause of any potential systematic variability in the transmission spectrum.
\begin{figure}
	\includegraphics[width=\linewidth]{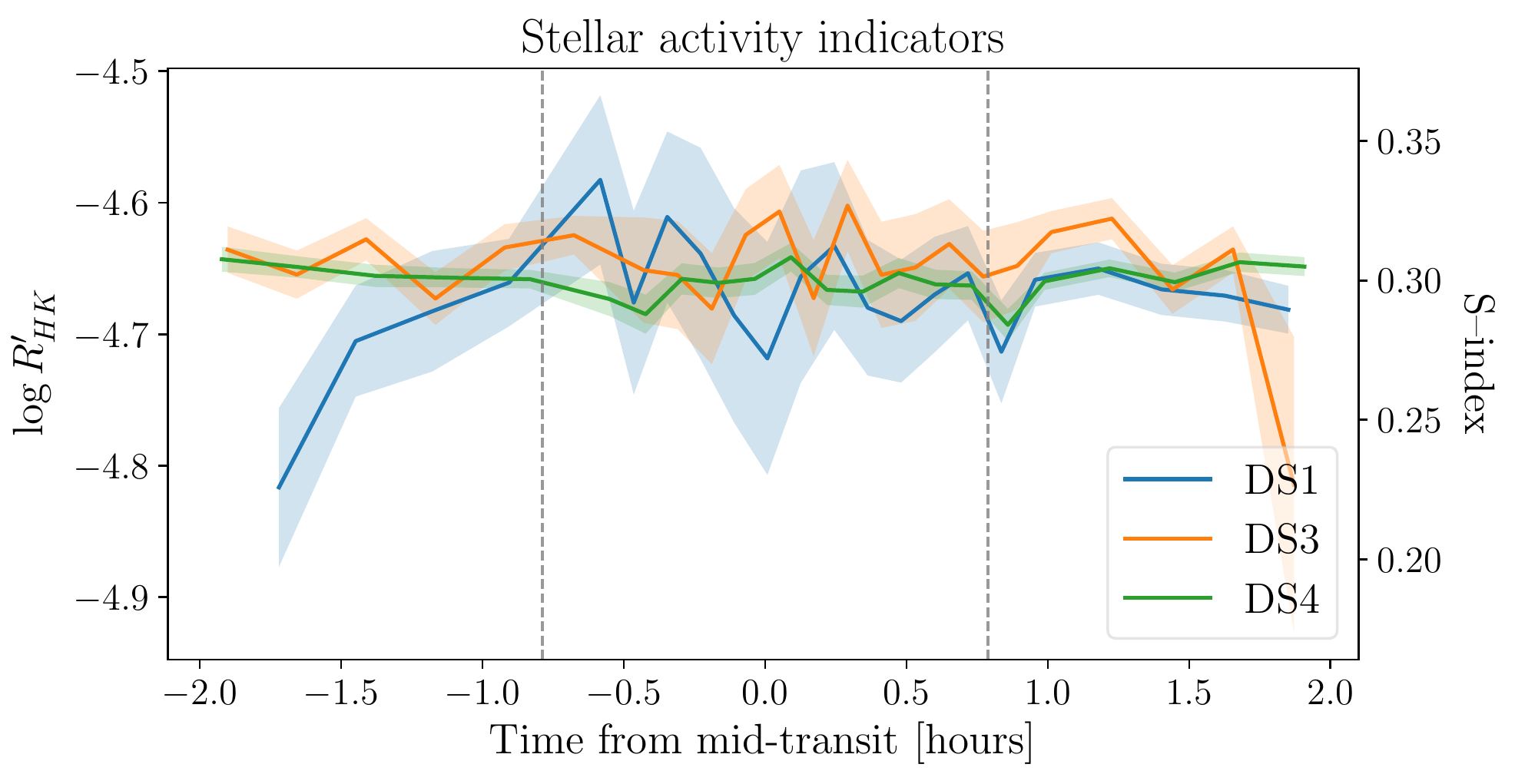}
    \caption{Variations of two stellar activity indicators measured with the ESPRESSO DAS for all exposures of the three data sets studied. The left y-axis shows the scale of variations in $\log R^{\prime}_{HK}$ indicative of only the chromospheric contribution, whereas the opposing axis represents the scaling for the S-index that includes contributions from both the photosphere and chromosphere, where the scale is calculated on the Mount Wilson scale \citep{Wilson1978}. The shaded regions present the standard errors and the vertical dashed lines indicate first and last transit contact points.}
    \label{fig:Activity}
\end{figure}
The chromospheric and photospheric activity levels of the star are probed through calculation of the S-index (both) and $\log R^{\prime}_{HK}$ (chromospheric only), which we additionally perform for all individual spectra, using the ESPRESSO Data Analysis Software\footnote{\url{ftp://ftp.eso.org/pub/dfs/pipelines/instruments/espresso-das/espda-pipeline-manual-1.2.0.pdf}} \citep[DAS v. 1.2.0;][]{Cupani2015}. Their variations throughout all four data sets are shown in the two panels of Fig. \ref{fig:Activity}, where both parameter values are indicative of moderate activity levels \citep{Gagne2016}. Additionally, we measure marginal variation in both parameters during DS1 observations, whereas the others are constant (DS3 \& DS4), or not determined precisely enough (DS2). Variations to activity levels can especially be problematic for transmission spectroscopy studies, as they can introduce differential effects that mimic exoplanetary atmospheric signals \citep{Oshagh2020}. This is of marginal concern here, as no detections have been made from the transmission spectrum.

\subsubsection{Validation of non-detections}
\label{sec:EMC}
We validate our non-detections of various lines by following the approach of \citet{Redfield2008} in performing a bootstrap analysis of the residual spectra, which is alternatively termed Empirical Monte Carlo (EMC). In a nutshell, this entails the calculation of three sets of transmission spectra, (1) the ``\textit{out-out}'' where only out of transit spectra are used to create the transmission spectrum, (2) the ``\textit{in-in}'' where only in-transit spectra are used, and (3) the ``\textit{in-out}'' which is the standard approach. For each of these, a subset of in and out of transit spectra are chosen at random and the entire analysis algorithm is run to create the final transmission spectrum. The binned depth in the core of the line being analyzed is measured, and this process is repeated many times to obtain a distribution of transmission signal depth. For a signal emanating from the exoplanetary atmosphere, the \textit{out-out} and \textit{in-in} distributions of line core depths are expected to be centered on zero, whereas the \textit{in-out} is expected at around the depth detected in the transmission spectrum. In all the line cores listed in Table \ref{tab:atomic upper limits}, we obtain Gaussian distributions centred on zero, for all three cases, validating those non-detections.

\subsection{Cross correlation analysis}
\label{section:CCF analysis}
\begin{figure*}
    \centering
    \includegraphics[width=\textwidth]{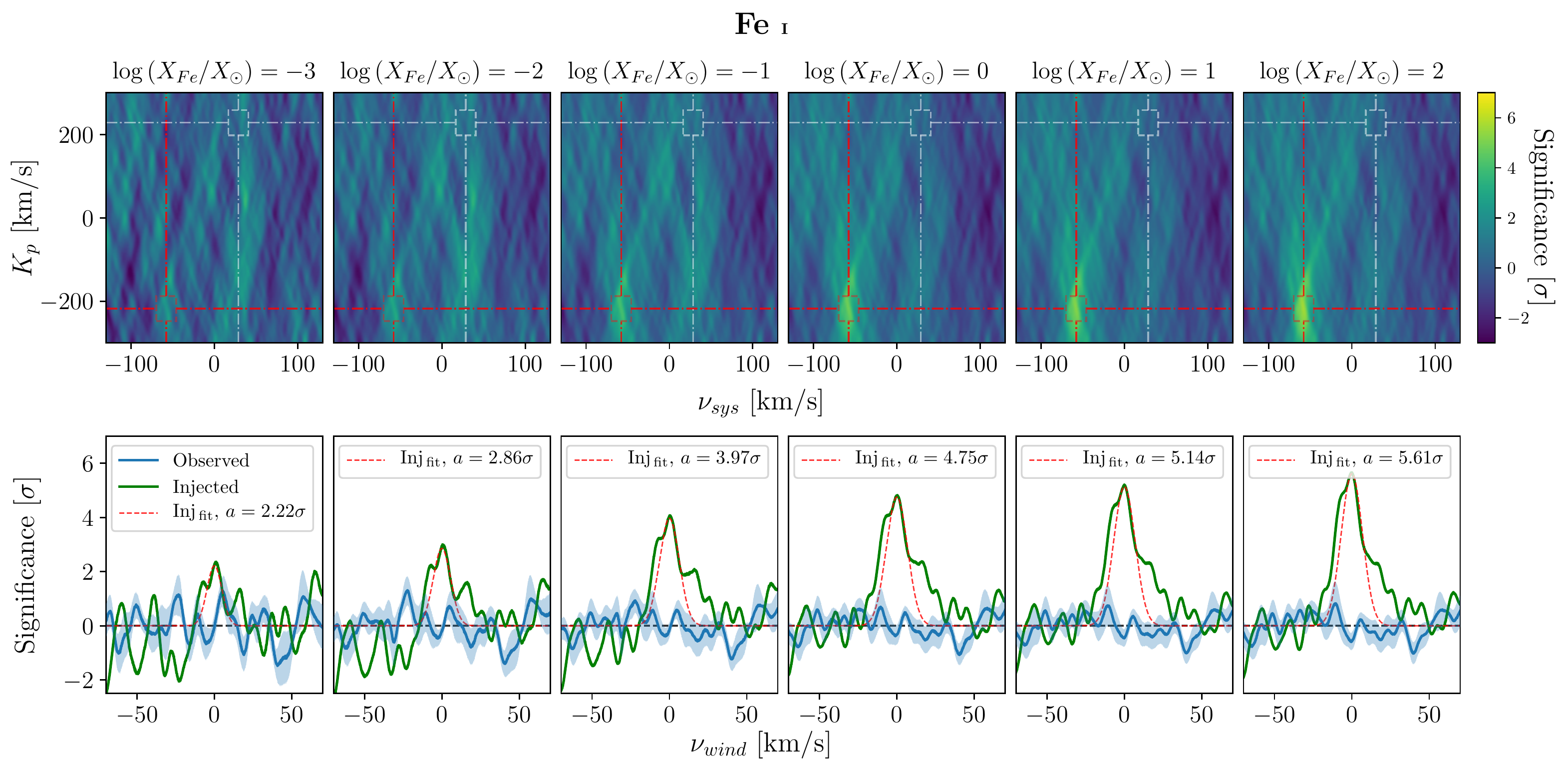}
    \caption{Cross correlation functions for Fe\,{\scriptsize I} co-added in velocity-velocity space (\textit{top}), as well as the planet rest-frame and the injected rest-frame signals (\textit{bottom}) both shifted to their respective zero points. The velocity-velocity maps each include an injected model, calculated for abundances relative to the solar value, given at the top of each panel. The white dashed-dot lines and boxes point to the expected location of a hypothetical planetary signal and their red equivalent to the injected signal. In the bottom panels the blue line is the signal at the $K_p$ of the planet, with the shaded region representing the 1$\sigma$ uncertainty. The green line is the same signal for $K_p$ at which the models were injected. The red dashed lines show the Gaussian fit to each injected peak to determine the significance at which each signal is recovered, with the amplitudes annotated in each panel.}
    \label{fig:Fe vel-vel}
\end{figure*}

\begin{figure*}
    \centering
    \includegraphics[width=\textwidth]{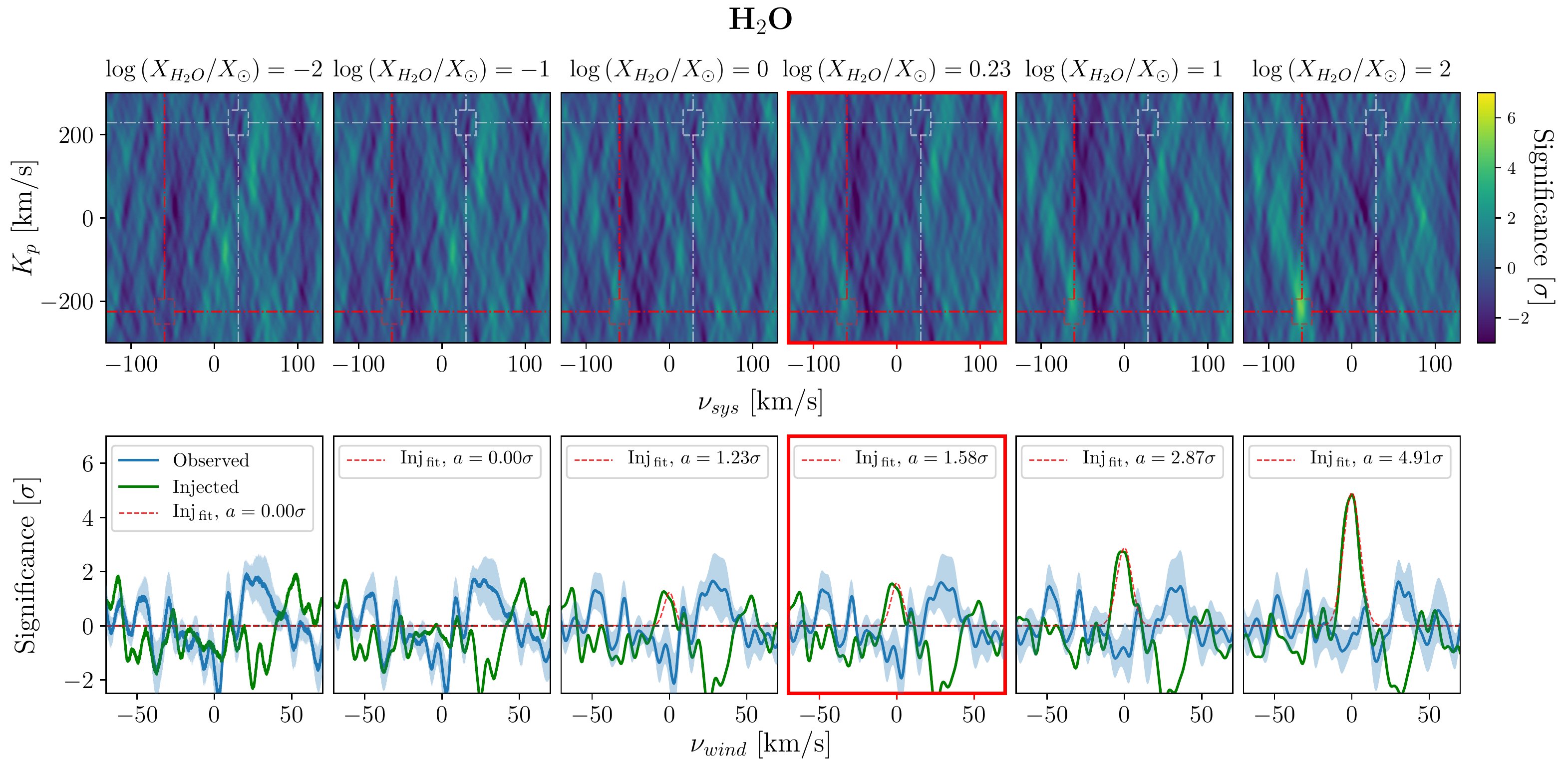}
    \caption{Same as Fig. \ref{fig:Fe vel-vel} but for H$_2$O template and injection models. The panels with red borders show the injected model representing the abundance value retrieved from the FORS2 low resolution transmission spectrum in \citet{Sedaghati2017}.}
    \label{fig:H2O vel-vel}
\end{figure*}

\begin{figure*}
    \centering
    \includegraphics[width=\textwidth]{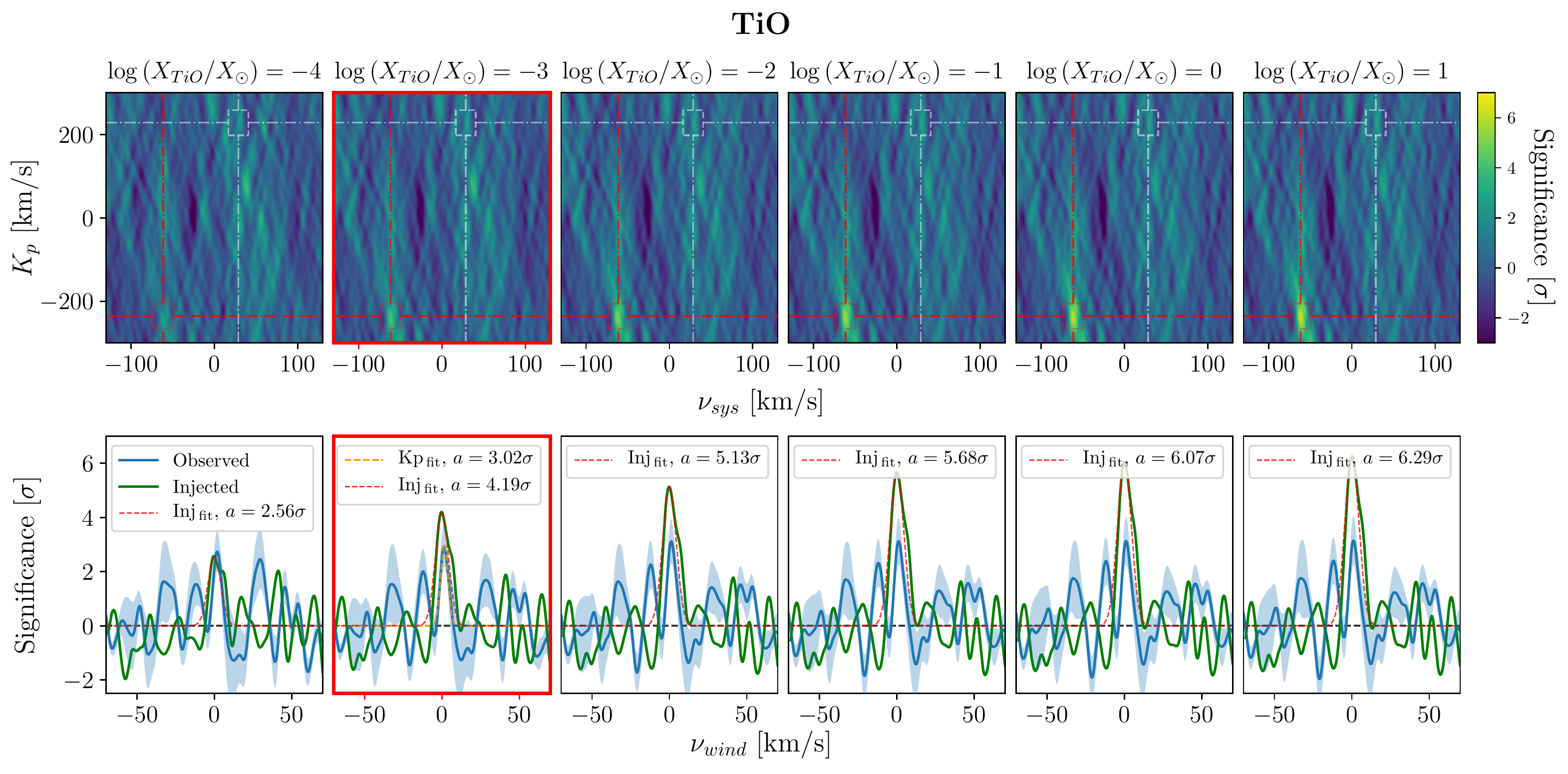}
    \caption{Same as Figs. \ref{fig:Fe vel-vel} and \ref{fig:H2O vel-vel} but for TiO models. Again the panels with red borders represent the injection models calculated for the abundance value retrieved from the FORS2 transmission spectrum. In the second panel of the bottom row, an additional Gaussian model is fitted to the apparent peak detected at the $K_p$ of planet (orange dashed line), with an amplitude of $3.02 \pm 0.15\,\sigma$, and centred on a radial velocity value $1.50 \pm 0.34$\,km/s offset from the calculated $\nu_{sys}$.}
    \label{fig:TiO vel-vel}
\end{figure*}

The cross correlation of a template, described in Section \ref{sec:CCF templates}, with the spectral series is performed for 
atmospheric models of Fe\,{\scriptsize I}, H$_2$O and TiO, in all cases sampling the RV space from $-$200 to $+$200\,km/s, at steps of 0.1\,km/s. In the calculation of $C(\nu,t)$ from equation \ref{eq:CCF}, we ignore from the sum those slices whose CCF's either show very low S/N ($\lesssim 1$) or present large systematic variations, e.g. the slices including telluric $O_2$ bands with saturated lines. This is a crucial step without which the S/N in the cross correlation maps would be significantly diminished. Each CCF is then normalized to its mean and a ``clean'' stellar CCF, derived from the weighted mean of all out of transit CCF's, is created. This star-only CCF is then divided out of all in-transit ones. This procedure is essentially equivalent to removing the stellar spectrum from the in-transit spectra and performing the CCF analysis on the residuals \citep[e.g][]{Snellen2010,deKok2013,Hoeijmakers2015,Brogi2016,Gibson2020}. We preferred this approach as a comparison between the maps produced by the two methods shows a marginally less noisy residual CCF map. This procedure produces a matrix of CCF's as described in equation \ref{eq:CCF}, where the rows are epochs of observations and the columns are the RV values at which the cross correlation is sampled. An example of such map is shown in the top panel of Fig. \ref{fig:CLV+RM}. The CLV+RM model, in velocity space, calculated from the cross correlation of simulated spectra (that include only the effect) with individual 
atmospheric models (c.f. middle panel of Fig. \ref{fig:CLV+RM}), is then subtracted from those aforementioned maps (via the application of a scaling factor, bottom panel of Fig. \ref{fig:CLV+RM}). The in-transit ``cleaned'' residual map is then shifted to a range of hypothetical $K_p$ rest-frames and collapsed along the RV axis to sample the CCF signal significance in $K_p$ space \citep{Snellen2010,Birkby2013}. This map, sometimes referred to as the ``velocity-velocity'' map, is divided by its standard deviation calculated from a large region where no signal is expected nor present, to calibrate it to its inherent noise level.

The above procedure is repeated for each species being probed, for spectral series that have had 
atmospheric models, calculated for a range of abundances spanning sub- to super solar values, injected into them. As {\tt petitRADTRANS} assumes mass fraction (MF; $X_i$), instead of volume mixing ratio (VMR; $n_i$) for abundances, we calculate the atmospheric models with individual metallicities relative to their corresponding solar mass fractions. The injected models, together with their corresponding abundances and the associated solar values are given in Table \ref{tab:model abundances}.

\begin{table}
    \caption{Range of 
    atmosphere models calculated for the purpose of injection and template creation.}
    \label{tab:model abundances}
    \begin{tabular}{lccc}
        \hline
         Species & Abundance         & Solar MF     & Reference \\
         & range ($\times\,X_\odot$) &  ($X_\odot$) & \\
         \hline \hline
         Fe\,{\scriptsize I} & 0.001 -- 100     & $1.6 \times 10^{-3}$ & \citet{Woitke2018}\\
         H$_2$O              & 0.01 -- 100     & $4.5 \times 10^{-3}$ & \citet{Woitke2018} \\
         TiO                 & 0.0001 -- 10,000 & $3.2 \times 10^{-6}$ & \citet{Woitke2018} \\
         \hline
    \end{tabular}
\end{table}

We chose to probe the three aforementioned species for two reasons. Fe\,{\scriptsize I} was searched for in order to more precisely determine the scaling necessary for removing the CLV+RM effects from the CCF maps, since there are plenty of neutral iron lines present in the stellar spectra. This means that the two effects are very clearly observed in the deformation of these many lines, which can be seen in the top panel of Fig. \ref{fig:CLV+RM}. Furthermore, it is complicated to detect the presence of Fe {\scriptsize I} from low resolution spectroscopy as it does not contain very strong singular transition lines, nor broad absorption bands, present for TiO or VO. Therefore, if present, previous transmission spectra of this exoplanetary atmosphere would have failed to detect it \citep{Sedaghati2017,Espinoza2019}. H$_2$O and TiO are probed since their presence was detected in previous studies \citep{Huitson2013,Sing2016,Sedaghati2017}, while other studies failed to confirm those detections \citep{Huitson2013,Espinoza2019}. Such discrepancies in low resolution spectroscopy are primarily attributed to the presence of stellar active regions on the stellar surface that introduce wavelength-dependent biases in the relative radius measurement of the planet \citep{Zellem2017,Rackham2017,Rackham2018,Rackham2019}. This point is further explored in section \ref{sec:Discussion}.

We performed the full cascade of cross correlation analysis for all four data sets. However, the first three data sets (DS1--DS3), observed before the upgrade of the ESPRESSO fiber link, are too noisy for any conclusions. Therefore, in this section we present results from the analysis of data only from DS4. Figs. \ref{fig:Fe vel-vel}, \ref{fig:H2O vel-vel} and \ref{fig:TiO vel-vel} show the CCF's co-added in velocity-velocity space for templates calculated for Fe\,{\scriptsize I}, H$_2$O and TiO, respectively. They simultaneously include injected models of each species, computed for a range of abundances spanning sub- to super-solar values, given in Table \ref{tab:model abundances}.

One very important caveat to note is that in calculating the atmospheric models used for both injection and CCF template creation, we assumed a fixed T/P profile. This profile is calculated using the formulation from \citet{Guillot2010}, assuming the equilibrium temperature from \citet{Sedaghati2017}. However, variations in the T/P profile also have an impact on the atmospheric model \citep[e.g.,][]{Hoeijmakers2020}, and subsequently would alter any conclusions presented in the following sections regarding upper limits of various atmospheric constituents. Therefore we caution the reader to take into account this caveat when interpreting any results presented below.

\begin{figure}
	\includegraphics[width=\linewidth]{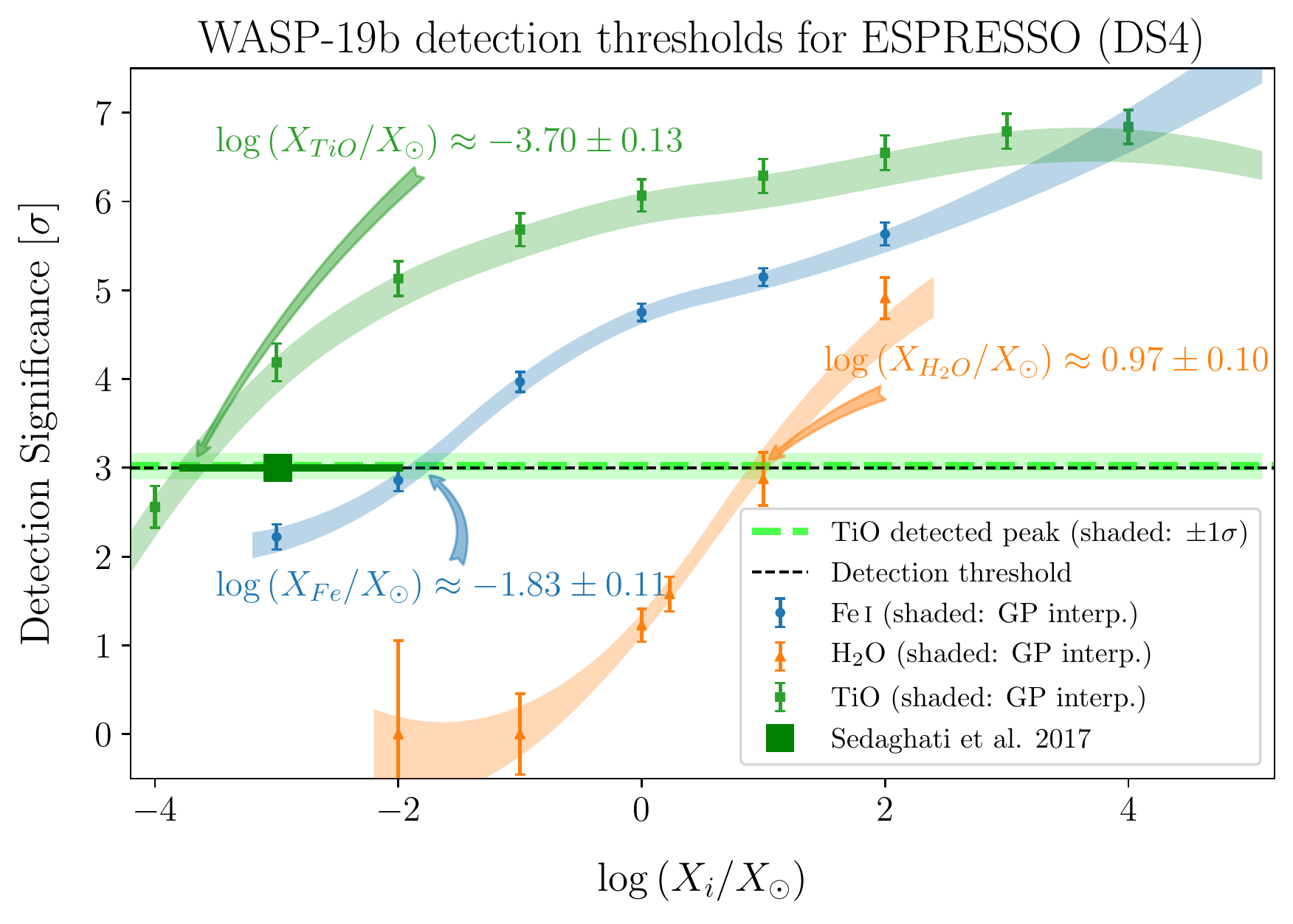}
    \caption{Significance values for the detected peaks of those injected 
    atmospheric models for Fe\,{\scriptsize I}, H$_2$O and TiO. The abundances with which the models are calculated are relative to their corresponding solar values ($X_\odot$). Each set of results is fitted with a non-parametric GP model to determine the detection threshold of the analysis for each species, which are shown as shaded regions. These hypothetical limits are annotated together with their uncertainties. The black dashed line is the optimistic 3$\sigma$ detection limit and the lime dashed line and the shaded region represent the peak of the detected TiO feature in the CCF map and its 1$\sigma$ uncertainty, respectively. The green square point presents the retrieved abundance of TiO by \citet{Sedaghati2017}, consistent with the detected peak in the ESPRESSO data.}
    \label{fig:Detection thresholds}
\end{figure}

\subsubsection*{Fe\,{\scriptsize I} non-detection}
We do not detect the presence of neutral iron for abundances down to the detection threshold of the observations. This is explored by the retrieval of the injected models shown in Fig. \ref{fig:Fe vel-vel} and summarized in Fig. \ref{fig:Detection thresholds}. Assuming an optimistic 3$\sigma$ detection threshold, retrieval of the injected signals shows that we can rule out the presence of Fe\,{\scriptsize I} in the upper atmosphere of WASP-19b down to sub-solar abundances, specifically $\log\,(X_{\textrm{Fe}}/X_\odot) \approx -1.83 \pm 0.11$. This is obtained through a non-parametric interpolation of the retrieved peaks (blue-shaded region in Fig. \ref{fig:Detection thresholds}), namely a Gaussian Process (GP) model with a squared exponential kernel, whose length scale is set to the distance between the individual data points.

\begin{figure}
	\includegraphics[width=\linewidth]{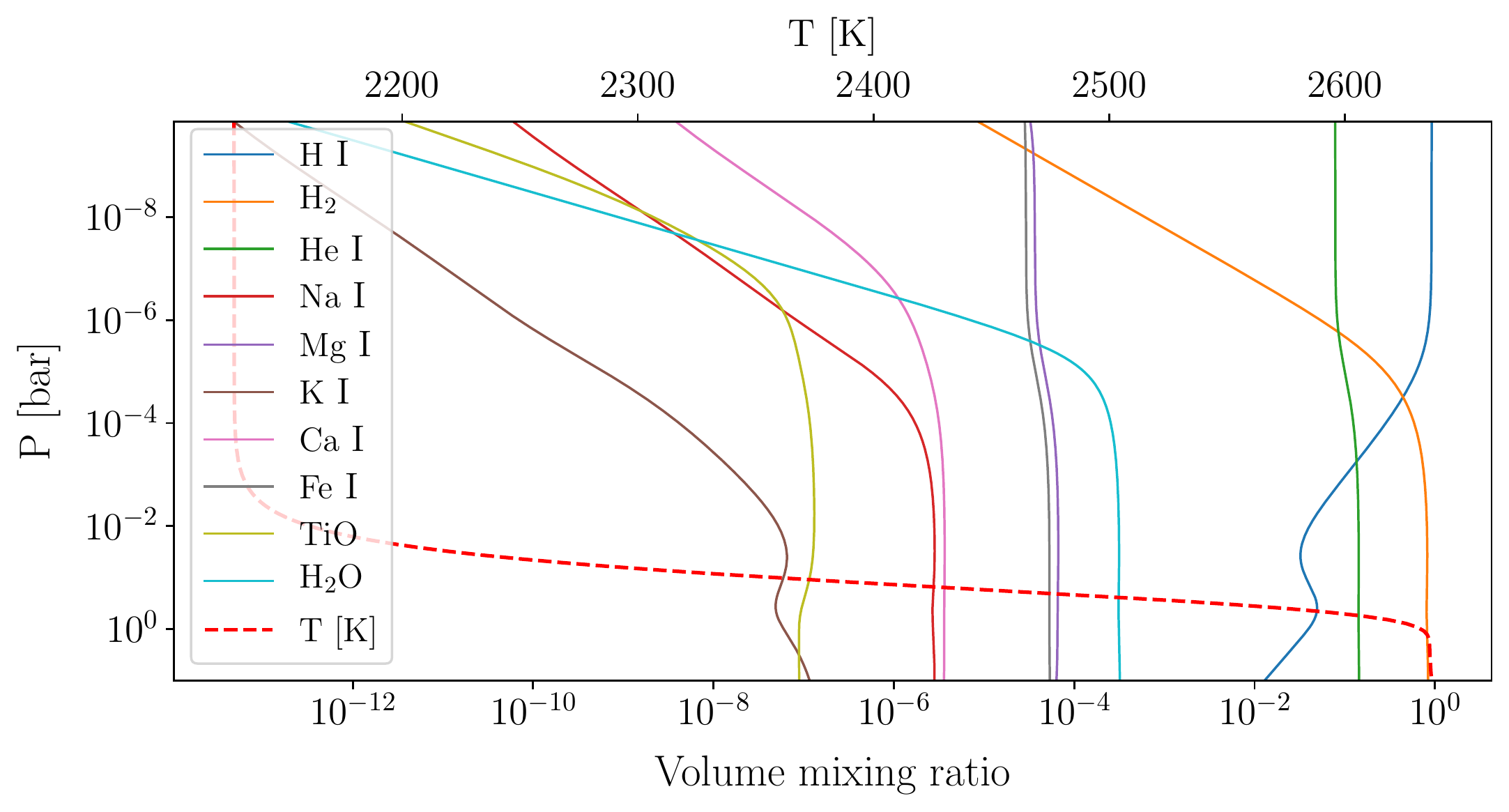}
    \caption{Abundance profiles of select species calculated for the \citet{Guillot2010} T-P profile presented as a red-dashed line, with its scale at the top. It is therefore deduced that Fe-bearing species do not make up a significant part of the total Fe inventory.}
    \label{fig:abundance profiles}
\end{figure}

Assuming chemical and hydrostatic equilibrium, with the P-T profile calculated via the analytical description of \citet{Guillot2010}, we compute abundance profiles for the species of interest, assuming solar photospheric metallicities for the elemental abundances. This calculation is performed with {\tt FastChem} \citep{Stock2018} with the profiles presented in Fig. \ref{fig:abundance profiles}. Fe\,{\scriptsize I} appears approximately constant with altitude, roughly equal to its solar abundance, meaning that Fe-bearing molecules do not make up a large part of the total Fe inventory. This conclusion could possibly be pointing to direct observational evidence of cold trapping processes hypothesised to deplete heavy metal abundances at the planet's terminator, as pointed out by \citet{Parmentier2013}. However, this also leaves open the possibility of depleting mechanisms stripping the protoplanetary disk of heavy metals.

\subsubsection*{Inability to detect H$_2$O}
The same procedure as above is also performed for 
atmospheric models of H$_2$O, where it is determined that its detection is only possible for large metallicities. Namely, the interpolation of detected peaks leads to a detection threshold of $\log\,(X_{\textrm{H}_2\textrm{O}}/X_\odot) \approx 0.97 \pm 0.10$, which means that H$_2$O is not detectable with this current experimental setup. \citet{Sedaghati2017} reported a slightly super-solar abundance of $\log\,(X_{\textrm{H}_2\textrm{O}}/X_\odot) \approx 0.23$ from the retrieval of FORS2 low resolution spectrum, and consequently we calculated an additional 
atmosphere model for this exact mass fraction and injected it into the spectra (red panels of Fig. \ref{fig:H2O vel-vel}). The cross correlation analysis shows that this abundance is far below the optimistic 3$\sigma$ threshold ($\sim$\,1.5$\sigma$). This conclusion is presented systematically in Fig. \ref{fig:Detection thresholds}.

It has previously been shown that H$_2$O is particularly difficult to detect using high dispersion spectroscopy operating in the visible regime \citep{Allart2017,Allart2020,Sanchez-Lopez2020}, such as ESPRESSO, where extremely high S/N spectra are required for the calculation of a reasonably precise CCF. The relatively faint host star, together with the short transit duration prevents us from obtaining such high precision spectra, even on an 8-m class telescope equipped with a state of the art high resolution spectrograph. The strength of a potential CCF signal can be improved upon by observing the transit in the near-IR with dedicated instrumentation. For instance multiple studies have reported the presence of H$_2$O in exoplanetary atmospheres using the near-IR arm of the CARMENES \citep{Quirrenbach2010} spectrograph \citep{Sanchez-Lopez2019,Alonso-Floriano2019}, as well as similar detections \citep{Brogi2018} with the GIANO \citep{Oliva2004,Carleo2020} spectrograph operating in the near-IR alongside HARPS-N at the TNG.

\subsubsection*{Possible hint of TiO}
Similarly, we probed the presence of TiO in the atmosphere of WASP-19b, whose presence was previously reported by \citet{Sedaghati2017} at sub-solar abundance of $\log X_{\textrm{TiO}} \approx -8.5$. Subsequently, we also included this specific mass fraction in the range of calculated models for the purpose of injection, which is shown as the red panels in Fig. \ref{fig:TiO vel-vel}. The same figure presents the cross correlation analysis of the data, including a variety of models calculated for a range of atmospheric TiO abundances, covering sub- to super-solar values (c.f. Table \ref{tab:model abundances}).

Similar to the previous sub-section, we determine TiO detection threshold of DS4 through a GP non-parametric interpolation of the detected peaks of the injected signals. As annotated in Fig. \ref{fig:Detection thresholds}, this limit is set to $\log\,(X_{\textrm{TiO}}/X_\odot) \approx -3.70 \pm 0.13$. As evident in all panels of Fig. \ref{fig:TiO vel-vel}, we note the presence of a peak at the calculated $K_{p}$ of the planet and close to the $\nu_{\textrm{sys}}$ of the system. Specifically, this peak is offset from the systemic velocity by $1.50 \pm 0.34$\,km/s. This could be attributed to a multitude of factors including, but not limited to, (1) dynamics in the upper atmosphere 
\citep[e.g.][]{Snellen2010,Kempton2012,Brogi2016,Ehrenreich2020,Seidel2020} although the shift has the wrong sign for day-to-night transport, and/or (2) imprecisions in the calculated wavelengths for TiO transitions, affecting models from which the CCF templates are produced \citep[e.g.][]{Hoeijmakers2015,Piette2020}. However, this peak is detected at an amplitude of 3.02\,$\pm$\,0.15\,$\sigma$, and subsequently \textit{not statistically significant}. It is therefore, at best, a possible hint for the presence of TiO in the upper atmosphere of WASP-19b.

This detected peak is consistent with the sub-solar abundance of TiO ($\log\,(X_{\textrm{TiO}}/X_\odot) = -2.99^{+1.03}_{-0.81}$, presented in Fig. \ref{fig:Detection thresholds} as a green square data point) in upper atmosphere of WASP-19b, retrieved by \citet{Sedaghati2017} from FORS2 low-resolution transmission spectroscopy. In section \ref{sec:Discussion} we now discuss possible paths to reconciliation and consistency between that detection and the reported non-detection by \citet{Espinoza2019}.

\section{Discussion}
\label{sec:Discussion}
Our results presented in the previous section have demonstrated the non-detection of a variety of neutral atomic species through narrow-band transmission spectroscopy of strong individual absorption transitional lines. Those include species such as atomic hydrogen, sodium, potassium, calcium and magnesium. Additionally, we probed the atmosphere for atomic and molecular species with a large number of relatively weaker absorption lines/bands, using the cross correlation technique in order to co-add any potential atmospheric signal from all possible transitions. This was done through the calculation of 
atmospheric models for each species and using them as cross correlation templates to sum any atmospheric transmitted signal in velocity space. Through this procedure we reported a non-detection of Fe\,{\scriptsize I} and the inability to detect the presence of H$_2$O due to inadequate S/N in the individual spectra. Finally, we detect the presence of a peak in the cross correlation map of TiO at the expected location of the planetary orbit, where again due to the low S/N in the data, the peak does not constitute a statistically significant detection. This therefore does not mean a confirmation of previous detection from low resolution spectroscopy, and merely hints at a possible presence.

We now search for further possible consistencies between this high resolution study and previously obtained transmission spectra through low resolution differential spectrophotometric studies, as well as discuss our findings in the context of all previous conclusions made about the presence of TiO in the atmosphere of WASP-19b. We shall also consider the impact of the presence of unocculted active regions on the stellar surface on those results.

\subsection{Chromatic Rossiter-McLaughlin (CRM)}

\begin{table}
    \caption{Transmission spectrum values obtained from modeling of the CRM effect, utilizing RV curves from DS4.}
    \centering
    \label{tab:CRM}
    \begin{tabular}{lcc}
        \hline
         Bin centre &   Bin width &    R$_p$/R$_\star$ \\
         $[$\AA$]$      &   $[$\AA$]$     &            \\
        \hline \hline
         3852.2  &   90.8      &  $ 0.1460 ^{+0.0241}_{-0.0246}$\\
         3977.3  &   95.6      &  $ 0.1440 ^{+0.0154}_{-0.0156}$\\
         4110.8  &   101.0     &  $ 0.1506 ^{+0.0096}_{-0.0097}$\\
         4253.6  &   107.0     &  $ 0.1507 ^{+0.0079}_{-0.0080}$\\
         4406.7  &   113.5     &  $ 0.1385 ^{+0.0084}_{-0.0084}$\\
         4571.2  &   120.8     &  $ 0.1451 ^{+0.0080}_{-0.0081}$\\
         4748.5  &   128.8     &  $ 0.1488 ^{+0.0060}_{-0.0060}$\\
         4940.1  &   137.7     &  $ 0.1328 ^{+0.0059}_{-0.0059}$\\
         5147.8  &   125.9     &  $ 0.1436 ^{+0.0060}_{-0.0060}$\\
         5328.8  &   156.7     &  $ 0.1543 ^{+0.0068}_{-0.0068}$\\
         5571.1  &   169.4     &  $ 0.1443 ^{+0.0071}_{-0.0070}$\\
         5848.2  &   213.8     &  $ 0.1514 ^{+0.0085}_{-0.0086}$\\
         6190.4  &   204.0     &  $ 0.1407 ^{+0.0067}_{-0.0067}$\\
         6562.8  &   261.4     &  $ 0.1408 ^{+0.0091}_{-0.0091}$\\
         7388.9  &   577.2     &  $ 0.1497 ^{+0.0093}_{-0.0092}$\\
         \hline
    \end{tabular}
\end{table}

\begin{figure}
	\includegraphics[width=\linewidth]{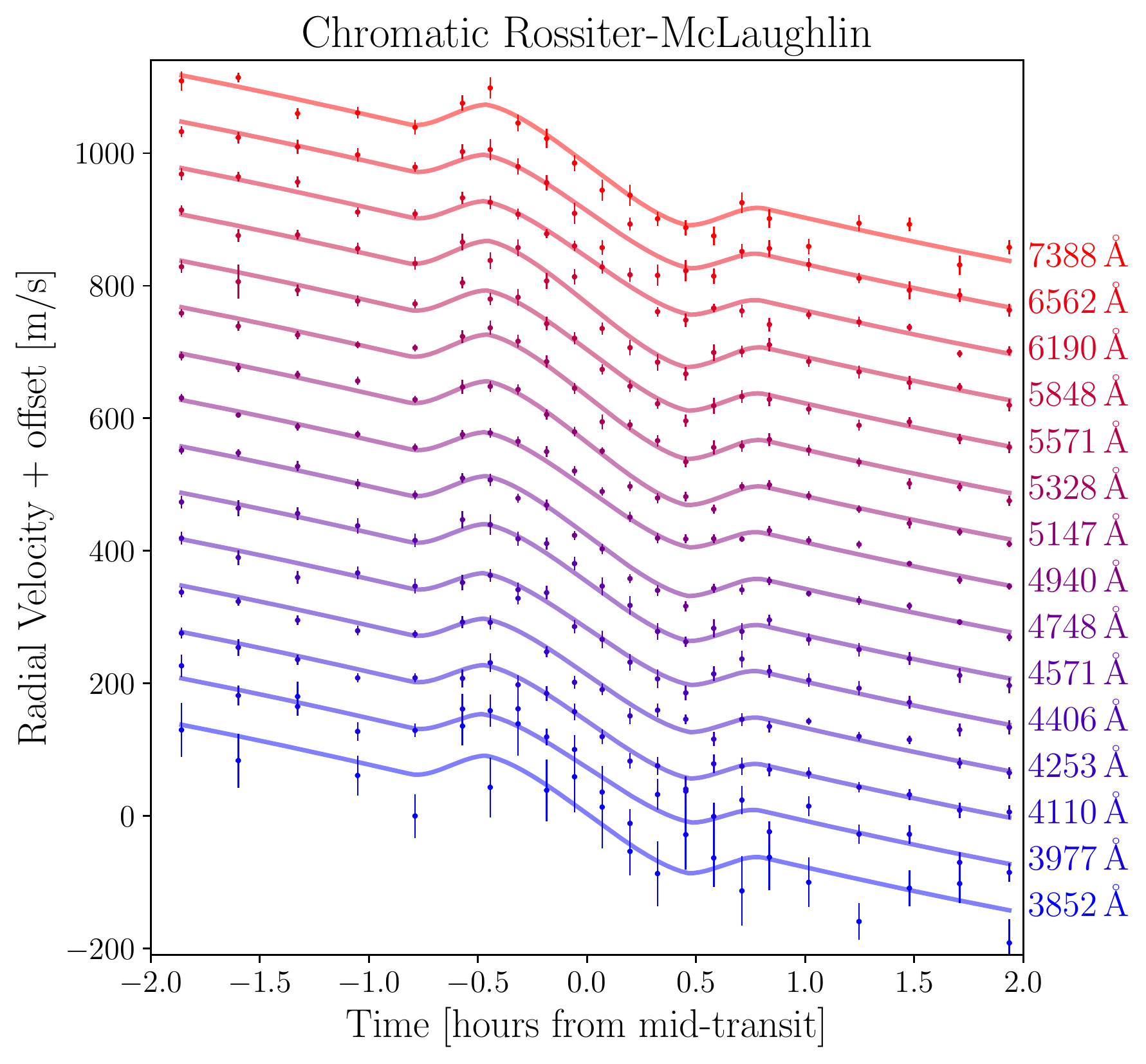}
    \caption{The chromatic Rossiter-McLaughlin measurements for the determination of planetary radius variation as a function of wavelength. The curves have been shifted by 100\,m/s for clarity. Each curve constitutes measurements from the sum of 5 spectral orders, the central wavelength of which is given to the right of each curve. Due to the non-linear nature of Echelle orders, as well as the discarded orders, these values are not evenly spaced. The solid lines present the best-fit solutions from the MCMC analysis. Please note that due to the scale of the y-axis, the error bars of the RV values are not distinguishable from the data points. These uncertainties have therefore been presented in the Appendix table \ref{tab:CRM RVs}.}
    \label{fig:Chromatic RM}
\end{figure}
A secondary channel through which an exoplanetary atmosphere could be probed from high resolution observations is the chromatic Rossiter-McLaughlin (CRM) measurement \citep[e.g.][]{DiGloria2015,Boldt2020,Oshagh2020,Palle2020,Santos2020}. This approach involves the determination of exoplanetary apparent radius variations as a function of wavelength, analogous to low resolution studies. To this end, one measures the RV variations during observations from individual spectral orders instead of the entire spectral range. Modeling each of these RV curves with an RM model, described previously in section \ref{sec:Star model}, results in the measurement of relative planetary radius as a function of mid-order wavelength.

To perform such analysis we utilized the order-by-order CCF produced by the ESPRESSO Data Reduction Software (DRS), which uses a synthetic G8 stellar binary mask 
identical to the spectral type of the target, as the template. For the analysis of the CRM effect we utilize the data from DS4 only, for the same reasons as the modelling of the white light RM effect, presented earlier in section \ref{sec:Star model}. The CCF's from all pairs of slices of each order were weight combined (85 spectral orders in total) of which 8 were discarded\footnote{The deleted orders are: [59, 68, 74 -- 78, 82], for which the DRS in fact does not calculate the CCF.} due to large systematic noise attributed to the presence of very deep or saturated telluric lines. We firstly determined the minimum number of spectral orders that needed to be binned in order to produce reasonably precise RM curves, similar to the systematic study performed by \citet{Santos2020}, and was set to 5 spectral orders (10 slices). All these combined CCF's were fitted with a Gaussian function, using a Levenberg-Marquardt algorithm to solve the minimization relation, whose mean value is the RV and its estimated error is taken from the covariance matrix. This is then repeated for all the bins, with the results in given in the Appendix table \ref{tab:CRM RVs}. The time series of these measurements constitute the CRM effect, which are presented in Fig. \ref{fig:Chromatic RM}. Each of these individual RM curves are then modelled using the same formulation presented in section \ref{sec:Star model}. For those non-wavelength parameters we 
fix their values to the results obtained from the broadband RM fit\footnote{We also allow the mid-transit time and the systemic velocity, $\nu_{\textrm{sys}}$ to vary within a narrow uniform prior centered on their respective broadband results.} and only allow the wavelength dependent parameters, namely the scaled planetary radius ($R_p/R_\star$) and the limb darkening coefficient, to vary under no prior assumptions. The optimal parameter values, together with their posterior probability distributions were then determined through MCMC simulations of 100,000 iterations. The best fit models for the individual RM curves are also over-plotted in Fig. \ref{fig:Chromatic RM}, with the determined transmission spectrum values given in table \ref{tab:CRM}.

In addition to those two colour-dependent parameters, we initially included a linear slope in the theoretical model (as was described in section \ref{sec:Star model}) in order to account for the chromatic impact of possible stellar heterogeneities \citep{Boldt2020}. However, there again was no statistical evidence present for the inclusion of such slope, and therefore it is concluded that out CRM curves are not affected by stellar activity. Subsequently, any possible conclusions made from the analysis of such curves cannot be attributed to spot or faculae on the stellar surface. 

The measured radii as a function of bin centres, i.e. the transmission spectrum, are plotted in all panels of Fig. \ref{fig:LR retrievals} together with the transmission spectra obtained from the FORS2 \citep{Sedaghati2016} and IMACS \citep{Espinoza2019} observations. Comparison between the three sets of results reveals a tentative (due to lower precision in R$_p$/R$_\star$) confirmation of the enhanced planetary size towards the near-UV wavelengths of up to $\sim$\,8 scale-heights that was detected in the FORS2 spectrum and in stark contrast with the IMACS results, indicative of scattering from hazes in a cloud-free atmosphere corresponding to a slope stronger than what is expected from Rayleigh scattering. However, the much lower precision and resolution of the CRM transmission spectrum in the mid-visible range, owing to the discarded orders, as well as the wider wavelength ranges of Echelle orders in that regime, means that no conclusions could be made about the presence of TiO from this approach. Following similar reasoning, these data points are not included in the new retrieval analyses presented in the following sub-sections.

\subsection{Reanalysis of low-resolution WASP-19b transmission spectra}
\begin{figure}
	\includegraphics[width=\linewidth]{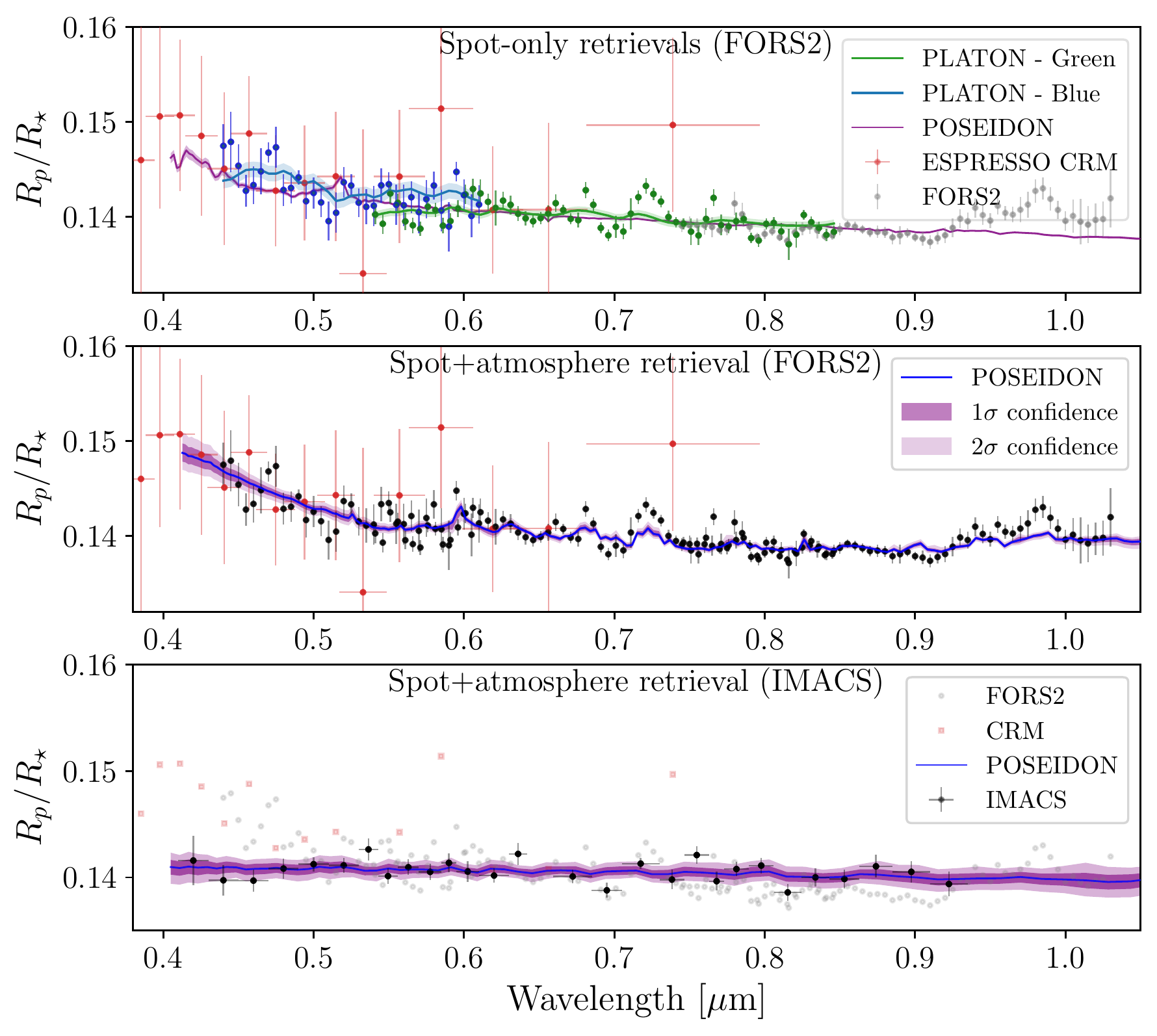}
    \caption{Retrieval analyses of WASP-19b broadband transmission spectra: \textit{(top)} WASP-19b transmission spectrum obtained with FORS2 \citep{Sedaghati2017} multi-epoch observations (Blue \& Green data sets color coded), as well as the spectrum obtained from CRM analysis of ESPRESSO spectra shown as red points. All spectral variations are assumed to arise from unocculted stellar heterogeneities and the planet is considered as an opaque disk of constant radius at all wavelengths (i.e. no atmosphere). The blue and green curves are the median stellar contamination-only fits from retrievals performed with {\tt PLATON} for the 600B (blue) and 600RI (green) data sets, respectively, while the purple curve is a similar retrieval with {\tt POSEIDON} run for the entire FORS2 data set with a strict prior on the photospheric temperature. The shaded regions represent 1-$\sigma$ confidence intervals for each retrieval. The results from the CRM analysis of ESPRESSO RV curves are overlaid for comparison as red points, but were not included in the retrieval analyses. \textit{(middle)} {\tt POSEIDON} retrieval of the FORS2 transmission spectrum where both a full exoplanetary atmosphere and unocculted stellar heterogeneities are considered. \textit{(bottom)} the same retrieval as the middle panel, but performed for the IMACS transmission spectrum from \citet{Espinoza2019}. FORS2 and ESPRESSO data are presented in the background for comparison.}
    \label{fig:LR retrievals}
\end{figure}

\citet{Espinoza2019} attributed discrepancies between their individual WASP-19b transmission spectra with IMACS to heterogeneities on the stellar surface in the form of spots or faculae, especially those unocculted by the transit chord of the planet. They suggested that such unocculted stellar heterogeneities could also explain the enhanced scattering slope and TiO spectral features present in the FORS2 spectrum of \citet{Sedaghati2017}. They further noted that the three constituent grism data sets comprising the FORS2 spectrum were likely impacted by different levels of stellar activity, due to differences in the epochs of observations. The FORS2 and IMACS spectra are compared in Fig. \ref{fig:LR retrievals}. In what follows, we present results from atmospheric retrieval analyses of the combined FORS2 data, the blue and green individual FORS2 data sets, and the IMACS data. 
Our goal is to assess the impact of stellar heterogeneities on the FORS2 spectrum, as well as searching for possible consistencies between the two seemingly discrepant results.


\subsubsection*{Activity-only retrievals of FORS2 spectra}

\begin{figure*}
    \centering
    \begin{subfigure}{0.3\linewidth}
         \centering
         \includegraphics[width=\linewidth]{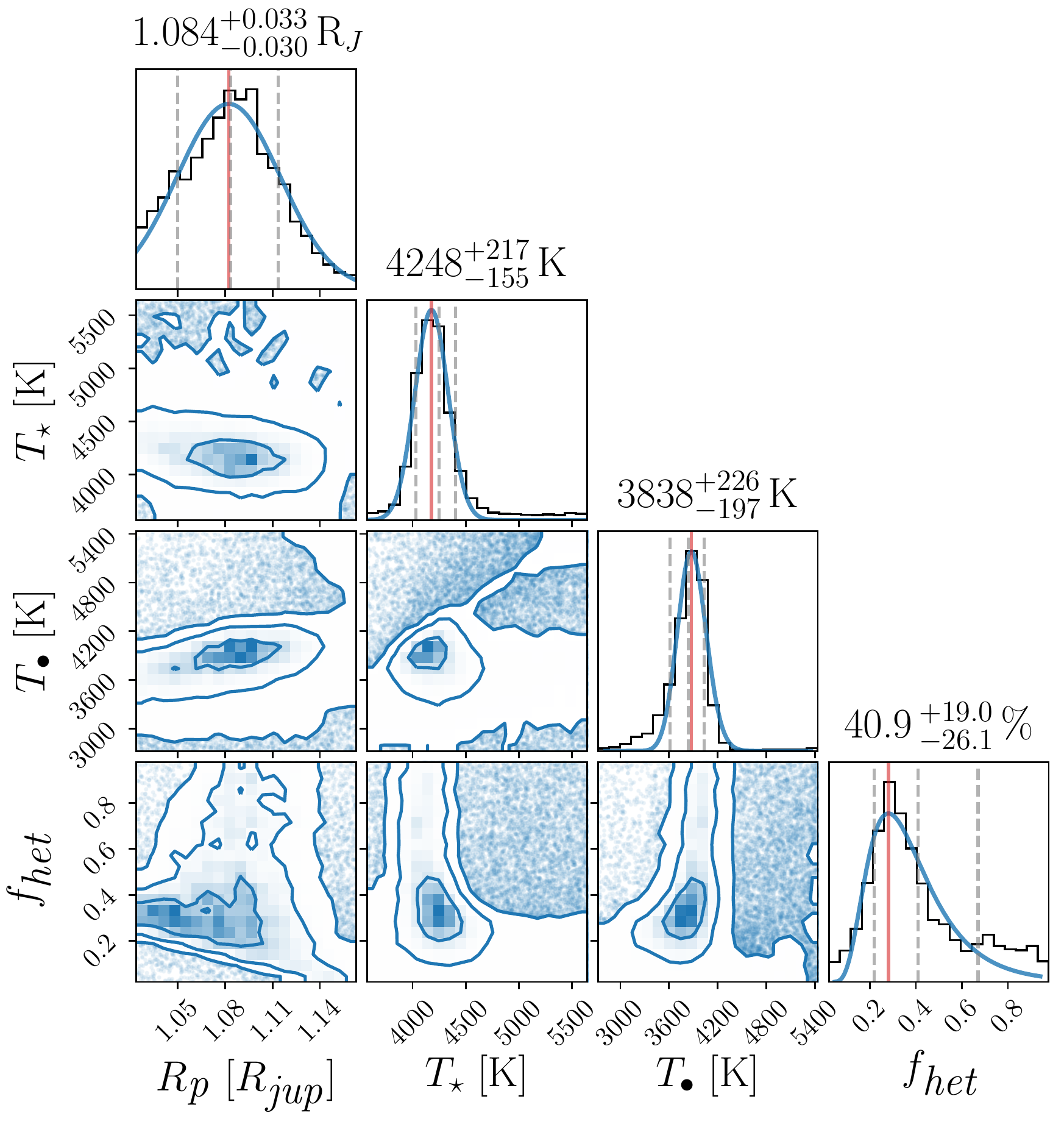}
         \caption{\centering {\tt PLATON} activity-only posteriors for FORS2 Blue}
         \label{fig:PLATON spot-only Blue}
     \end{subfigure}
    \begin{subfigure}{0.3\linewidth}
         \centering
         \includegraphics[width=\linewidth]{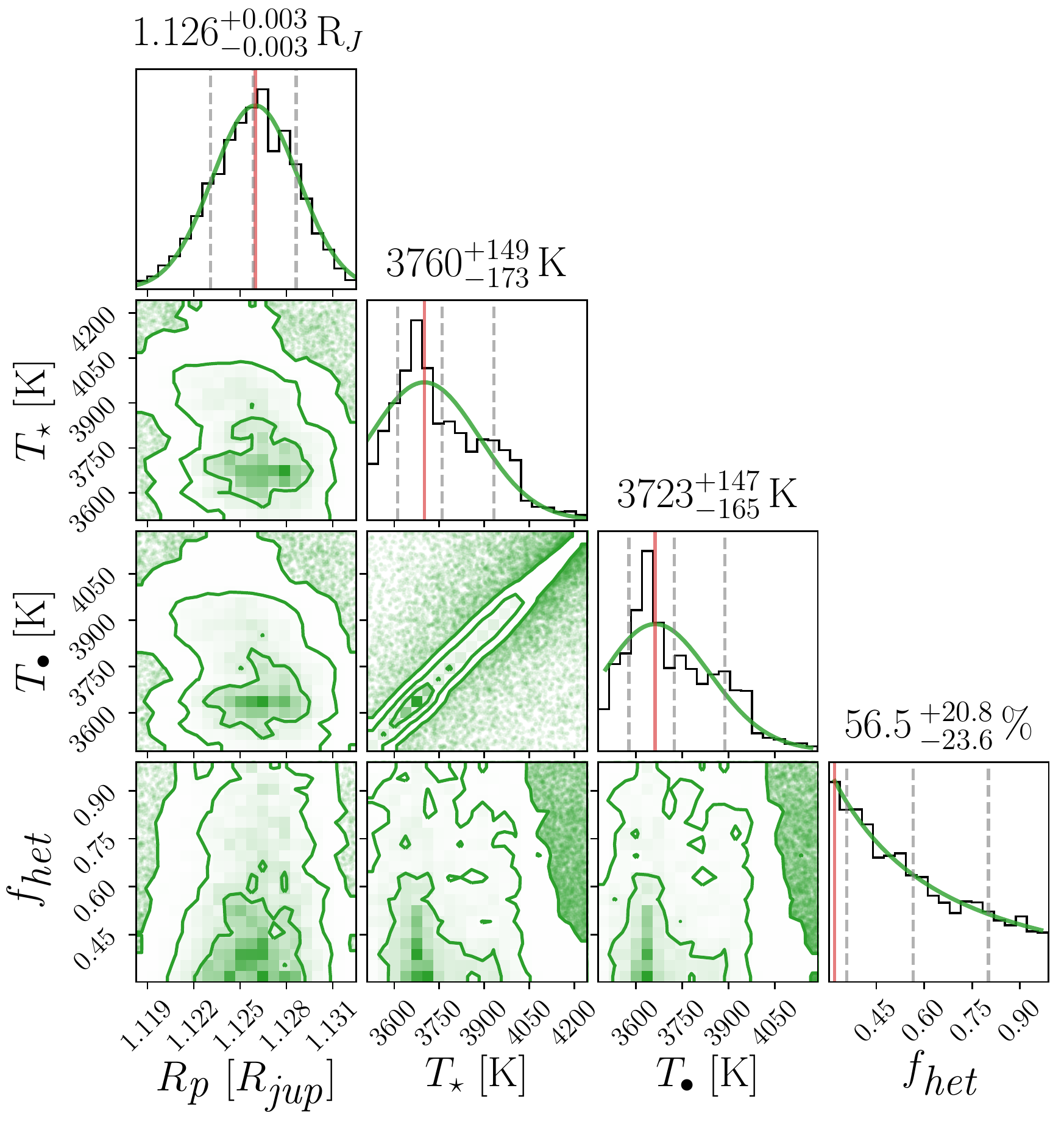}
         \caption{\centering {\tt PLATON} activity-only posteriors for FORS2 Green}
         \label{fig:PLATON spot-only Green}
     \end{subfigure}
    \begin{subfigure}{0.3\linewidth}
         \centering
         \includegraphics[width=\linewidth]{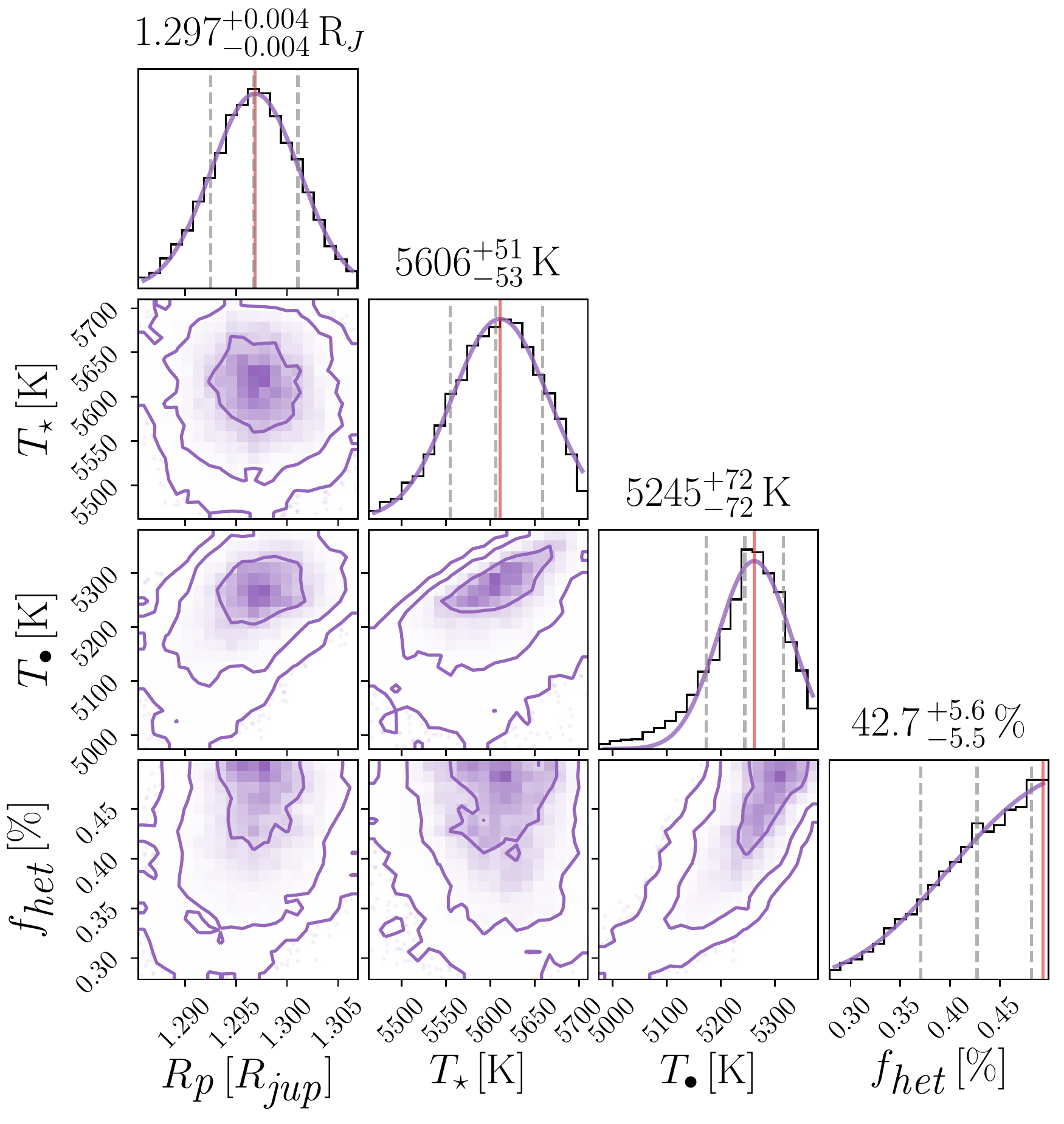}
         \caption{\centering  {\tt POSEIDON} activity-only posteriors for FORS2 All}
         \label{fig:POSEIDON FORS2 spot-only}
     \end{subfigure}

    \begin{subfigure}{0.45\linewidth}
         \centering
         \includegraphics[width=\linewidth]{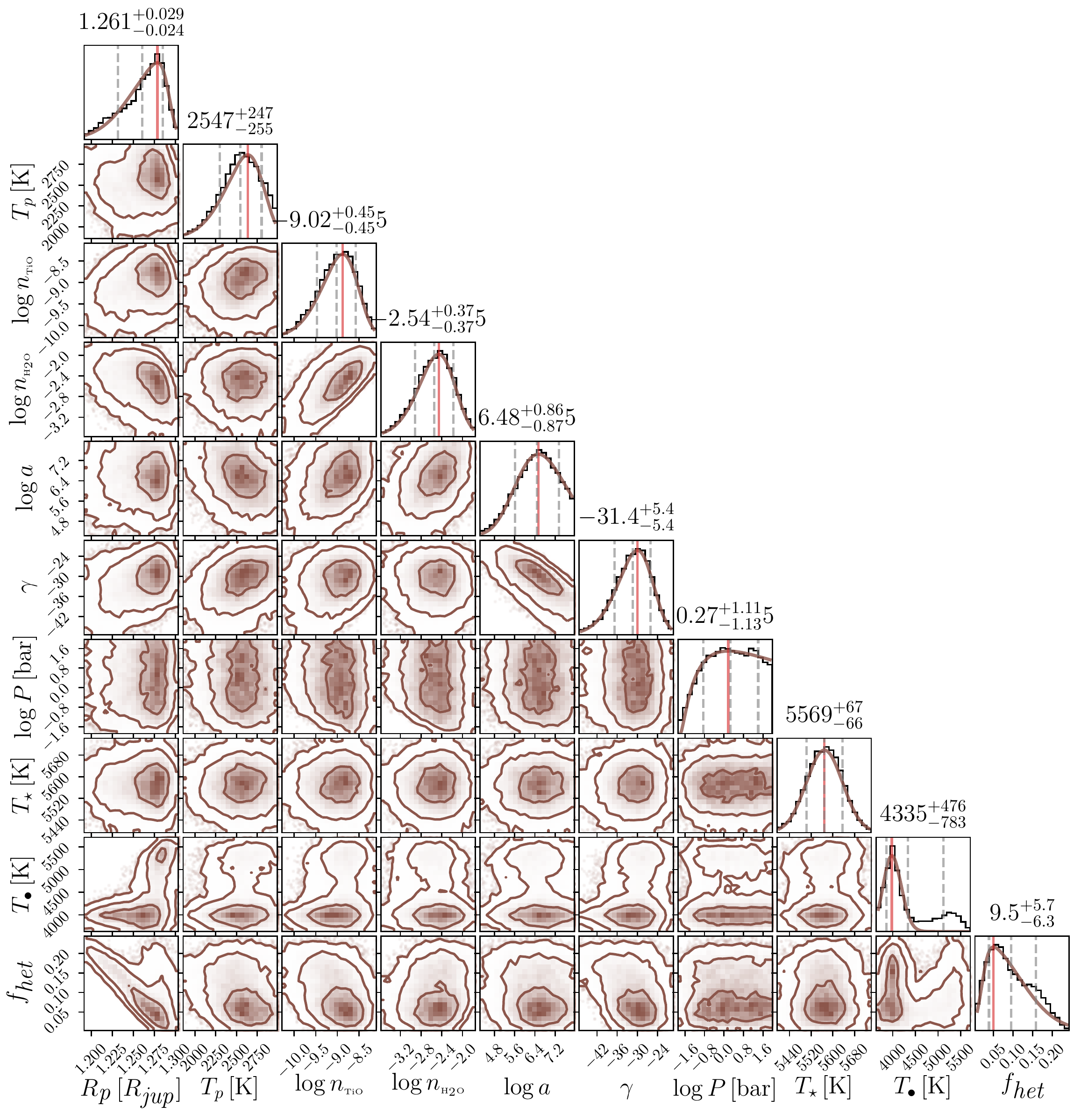}
         \caption{\centering {\tt POSEIDON} activity+atmosphere posteriors for FORS2}
         \label{fig:POSEIDON FORS2 spot-atmo}
     \end{subfigure}
    \begin{subfigure}{0.45\linewidth}
         \centering
         \includegraphics[width=\linewidth]{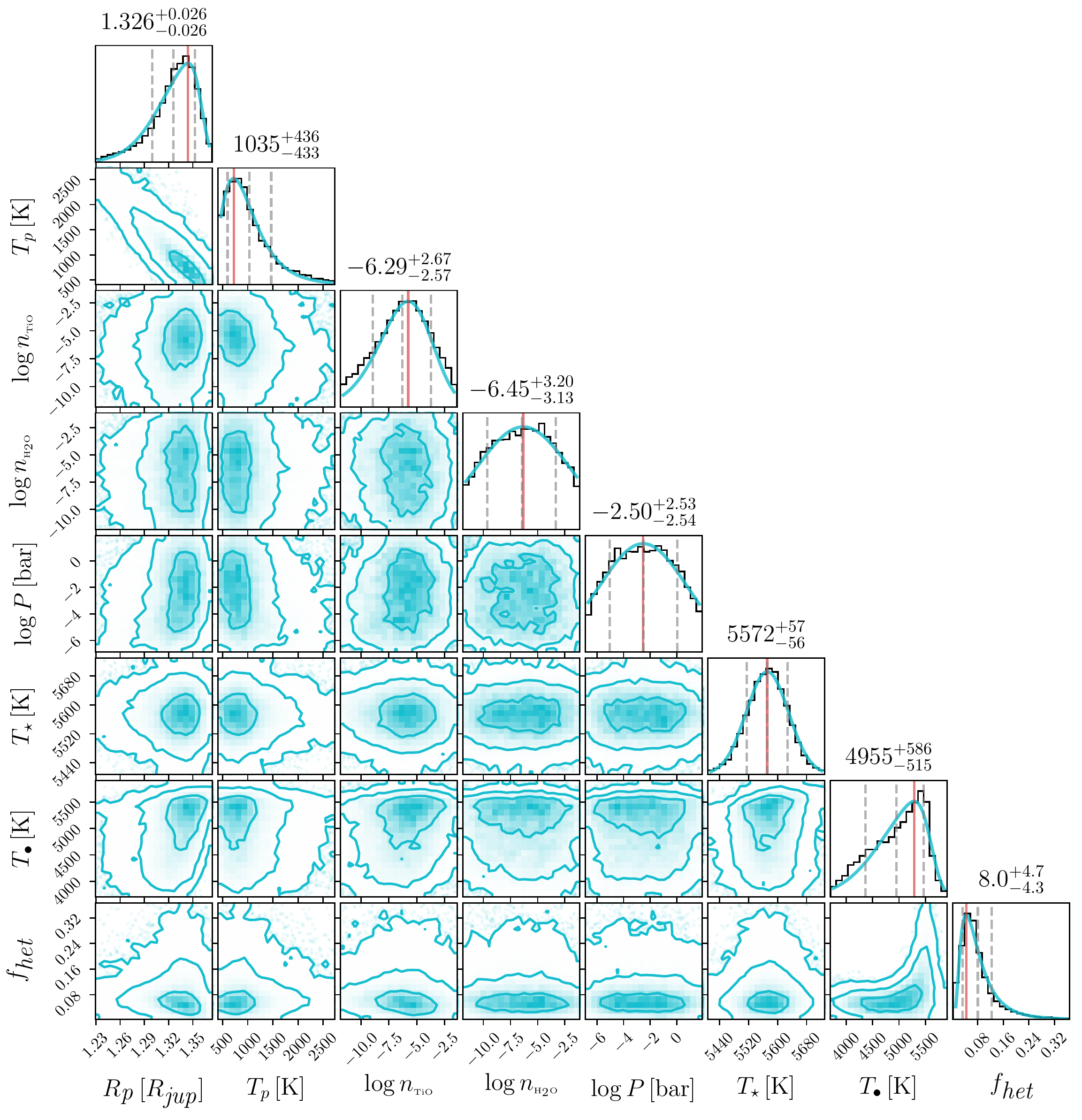}
         \caption{\centering {\tt POSEIDON} activity+atmosphere posteriors for IMACS}
         \label{fig:POSEIDON IMACS spot-atmo}
     \end{subfigure}

        \caption{Posterior probability distributions from retrieval analyses of FORS2 \& IMACS transmission spectra with {\tt PLATON} \& {\tt POSEIDON}, assuming both unocculted heterogeneity-only (top row) and atmosphere plus activity (bottom row) scenarios as sources of transit depth variations. \textit{(a)} {\tt PLATON} posterior for the activity-only fit to the Blue data set (600B grism) of the FORS2 spectrum with no prior assumed for the photospheric temperature; \textit{(b)} same as panel \textit{a} but for the Green data set (600RI) of the FORS2 spectrum; \textit{(c)} {\tt POSEIDON} activity-only posterior for the retrieval of the entire FORS2 spectrum, assuming a Gaussian prior on the stellar photospheric temperature; \textit{(d)} {\tt POSEIDON} posteriors for the activity plus atmosphere retrieval of the entire FORS2 spectrum, where abundances are given as log$_{10}$ volume mixing ratios and the scattering is parametrised through the relation $\sigma(\lambda) = a\,\sigma_0 (\lambda / \lambda_0)^\gamma$, with $\lambda_0$ a reference wavelength (350\,nm) and $\sigma_0$ the H$_2$-Rayleigh scattering cross-section at the reference wavelength (5.31\,$\times$\,10$^{-31}$\,m$^2$); and \textit{(e)} the same as panel \textit{d} but for the IMACS data set, where no scattering is detected. All distributions are derived from MultiNest runs with 2000 live points. The mean and 1$\sigma$ quartiles are represented by grey dashed lines and are annotated above each histogram. The most probable values of all distributions are indicated with solid red lines, calculated through fitting either skewed-normal or log-normal functions.}
        \label{fig:Posteriors}
\end{figure*}

\begin{table*}
 \caption{Summary of results from retrievals of low resolution transmission spectra of \citet{Sedaghati2017} and \citet{Espinoza2019}.}
 \label{tab:retrieval results}
 \begin{tabular}{lccccc}
  \hline 
    & \multicolumn{2}{c}{{\tt PLATON}} & \multicolumn{3}{c}{{\tt POSEIDON}} \\
  \cline{2-6}
  Parameter & \multicolumn{2}{c}{Spot-only} & Spot-only & \multicolumn{2}{c}{Spot+Atmosphere} \\
  \cline{2-6}
   & FORS2 (Blue) & FORS2 (Green) &  FORS2 & FORS2 & IMACS \\  
  \hline \hline
  T$_\star$ [K]                            & 4248$^{+217}_{-155}$       & 3760$^{+149}_{-173}$      & 5603$^{+54}_{-55}$        & 5569$^{+67}_{-66}$        & 5572$^{+57}_{-56}$ \\ 
  T$_\bullet$ [K]                          & 3838$^{+226}_{-197}$       & 3723$^{+147}_{-165}$      & 5215$^{+68}_{-99}$        & 4335$^{+476}_{-783}$      & 4955$^{+586}_{-515}$\\ 
  $f_{\textrm{\textit{het}}}$ [\%]         & 40.9$^{+19.0}_{-26.1}$     & 56.5$^{+20.8}_{-23.6}$    & 41.7$^{+6.3}_{-6.4}$      & 9.5$^{+5.7}_{-6.3}$       & 8.0$^{+4.7}_{-4.3}$\\ 
  R$_p$ [R$_{\textrm{\textit{jup}}}$]      & 1.084$^{+0.033}_{-0.030}$  & 1.126\,$\pm$\,0.003       & 1.296$^{+0.004}_{-0.005}$ & 1.261$^{+0.029}_{-0.024}$ & 1.326\,$\pm$\,0.026 \\
  T$_p$ [K]                                & --                         & --                        & --                        & 2547$^{+247}_{-255}$      & 1035$^{+436}_{-433}$      \\
  $\log \textrm{P}_{\textrm{cloud}}$ [bar] & <$-4$>                     & <$-4$>                    & --                    & 0.27$^{+1.11}_{-1.13}$    & $-$2.50$^{+2.53}_{-2.54}$ \\ 
  $\log n_{\textrm{{\tiny TiO}}}$\,$^a$                  & --                         & --                        & --                        & $-$9.02\,$\pm$0.45        & $-$6.29$^{+2.67}_{-2.57}$\\
  $\log n_{ \textrm{{\tiny H}}_2\textrm{{\tiny O}}}$\,$^a$        & --                         & --                        & --                        & $-$2.54\,$\pm$0.37        & $-$6.45$^{+3.20}_{-3.13}$ \\
  \hline
  \multicolumn{6}{l}{$^a$ Abundances are given as $\log\,(n_i)$, i.e. Volume Mixing Ratio (VMR), as opposed to the Mass Fraction}\\
  \multicolumn{6}{l}{~~~(MF) units utilized in Section \ref{sec:Results}, where $X_i \approx \frac{\mu_i}{\mu_{\textrm{atm}}} n_i$, with $\mu$ being the mean atomic/molecular weights.}  
 \end{tabular}
\end{table*}

To assess the concern of stellar heterogeneity, be it spots or faculae, being solely responsible for the signature of TiO and the enhanced scattering slope towards the blue end of the FORS2 spectrum, we ran a series of atmospheric retrievals under the assumption that the planet has a constant radius at all wavelengths, implying that it does not contain an atmosphere. 
This is done for individual epoch data sets (from which TiO and scattering was detected), as well as the combined transmission spectrum.

We first ran retrievals using the {\tt PLATON} package \citep{Zhang2019,Zhang2020}, employing nested sampling \citep{Skilling2004} via the {\tt dynesty} code \citep{Speagle2020} to explore the parameter space with 2000 live points. 
In these activity-only runs 
we set an uninformative uniform prior for the stellar effective temperature ($T_\star$), covering spectral types down to M, a surface heterogeneity filling factor up to 100\% of the stellar surface ($f_{\textrm{\textit{het}}}$), and allowed heterogeneity-photosphere temperature differences of up to 4000\,K ($| T_\star - T_\bullet |$). Additionally, we assumed the constant planetary radius as a free parameter due to uncertainties in the exact altitude of the reference pressure (the cloud top) and allow it to vary from 0.5 to 1.5 times the median radius from the spectrum. 

We fitted separately two individual data sets comprising the FORS2 spectrum; the Blue (600B grism) and the Green (600RI grism) sets. The fitted models from these analyses are both presented in the top panel of Fig. \ref{fig:LR retrievals}, where it is evident that unocculted spots or faculae alone cannot explain the planetary radius variations measured from FORS2 observations. However, certain spectral characteristics, such as an enhanced slope for the Blue-only data, can be captured by retrievals considering unocculted stellar spots. 

The fitted models 
converge on stellar effective temperatures of 4248$^{+217}_{-155}$\,K and 3760$^{+149}_{-173}$\,K for the Blue and Green data sets, respectively. These values are significantly below what has been measured for the star from different spectral 
analysis (5497\,$\pm$70\,K this study; 5500\,$\pm$\,100\,K \citet{Hebb2010}, 5568\,$\pm$\,71\,K \citet{Torres2012}). 
This is due to the fact that only cooler stars are expected to show strong TiO absorption lines in their spectra \citep{Bell1989,Keenan1989,O'Neal2004}. 
Values for the other fitted parameters are presented in posterior plots of these two runs (Figs. \ref{fig:PLATON spot-only Blue} \& \ref{fig:PLATON spot-only Green}), just within bounds of typically expected values for active low-mass stars \citep{Jackson2013}. This is in agreement with previous observations of large active regions for this star, measured as anomalies in transit light curves \citep{Huitson2013,Sedaghati2017,Espinoza2019}. The results from these two {\tt PLATON} retrievals are summarised in Table \ref{tab:retrieval results}.

Additionally, as a sanity check, we performed a further retrieval with the {\tt POSEIDON} atmospheric retrieval code \citep{MacDonald2017}, but this time set a physically motivated prior on the stellar photospheric temperature. This presents a more realistic and feasible solution to the question of unocculted stellar heterogeneity induced planetary radius variations. However, we note that this retrieval fits the entire FORS2 spectrum, which comprises of three different epochs of observations, while each of the individual data sets could potentially have been affected by different levels of activity. For this retrieval we set a Gaussian prior on the stellar photosphere temperature ($T_\star \in \mathcal{N}(5568,71)$\,K; \citet{Torres2012}). 
The best-fit model from this run is presented in the top panel of Fig. \ref{fig:LR retrievals} (purple curve), with the posteriors shown in Fig. \ref{fig:POSEIDON FORS2 spot-only} that includes the best-fit parameter values. 
Compared to a constant radius fiducial model, 
the presence of unocculted spots is significantly favoured with a Bayes factor \citep{Kass1993,Kass1995} $\ln{\mathcal{B}} = 113 $. However, it is evident in Fig. \ref{fig:LR retrievals} that this model fails to capture the spectral variations in the mid-visible range. Consequently, this activity-only model is disfavoured when compared to models including atmospheric TiO absorption, as we will now discuss.


\subsubsection*{Activity+atmosphere retrieval of FORS2 \& IMACS spectra}


Our simultaneous atmosphere and stellar heterogeneity retrievals were performed with {\tt POSEIDON}. Compared to the analysis in \citet{Sedaghati2017}, we made two important improvements to the retrieval methodology: (i) the inclusion of parametrised stellar heterogeneity (via the prescription in \citet{Pinhas2018}); and (ii) use of the ToTo ExoMol TiO line list \citep{McKemmish2019}, not available during the previous study. Similar to \citet{Sedaghati2017}, we considered a range of potential chemical species, temperatures, and cloud/haze properties, including absorption from Na, K, TiO, VO, CrH, FeH, H$_2$O, haze, clouds.

We detect the presence of TiO, H$_2$O, and a strong scattering haze, even when unocculted heterogeneities are included in the FORS2 retrieval analysis. This solution is conclusively favoured over our spot-only {\tt POSEIDON} retrieval with a Bayes factor of $\ln{\mathcal{B}} = 60$ (see Table \ref{tab:model statistics} for a summary of our Bayesian model comparisons). This best fitting model is plotted in the middle panel of Fig. \ref{fig:LR retrievals}, with its posterior distribution shown in Fig. \ref{fig:POSEIDON FORS2 spot-atmo}. When an atmosphere is included, we obtain a spot filling factor of 9.5\,$^{+5.7}_{-6.3}$\,\% and T$_\bullet$\,=\,4335\,$^{+476}_{-783}$\,K - approximately 1200\,K below the photospheric temperature (see Table \ref{tab:retrieval results}). The T$_\bullet$ posterior allows a wide range of solutions, permitting both cool unocculted spots and spots consistent with the photospheric temperature. For this reason, the Bayesian evidence prefers simpler models without the inclusion of unocculted heterogeneities (Table \ref{tab:model statistics}). 
\textit{The inclusion of unocculted spots in the retrieval model does not have a statistically significant impact on our retrieved abundances of TiO and H$_2$O, nor does it affect significantly the haze Rayleigh enhancement factor ($a$) or the scattering slope ($\gamma$).} 
\citet{Sedaghati2017}. However, our improved retrieval analysis results in two changes to TiO inferences from the FORS2 spectrum compared to \citet{Sedaghati2017}: (i) the TiO detection significance is revised from 7.7$\sigma$ to 4.7$\sigma$; and (ii) the retrieved TiO volume mixing ratio is around 10$\times$ greater ($\log n_{\textrm{{\tiny TiO}}} = -10 \pm 1 \rightarrow -9.02 \pm 0.45$). The improvement in the TiO abundance precision is an encouraging consequence of the new TiO line list from \citet{McKemmish2019}. In summary, we conclude that the presence of unocculted heterogeneities on the stellar surface cannot solely explain the previously obtained FORS2 spectrum. 



Finally, 
we also fitted the combined IMACS 
spectrum from \citet{Espinoza2019} for an activity and atmosphere model with {\tt POSEIDON}. 
Since \citet{Espinoza2019} only provided mean-subtracted transit depths for the IMACS spectrum, we added the white light $(R_p / R_\star)^2$ 
to the data for these retrievals. The median retrieved spectrum and confidence regions are plotted in the bottom panel of Fig. \ref{fig:LR retrievals}, with the corresponding posterior distribution shown in Fig. \ref{fig:POSEIDON IMACS spot-atmo}. 
These IMACS retrievals find no evidence for the presence of scattering haze, contrary to the FORS2 and ESPRESSO results. We find active region parameters consistent with what was determined by \citet{Espinoza2019}.  
However, we find that the overall best fitting model is a 4-parameter atmosphere-only model including TiO and VO (defined by T$_p$, R$_p$, $\log n_{\textrm{{\tiny TiO}}}$, and $\log n_{\textrm{{\tiny VO}}}$)\footnote{The retrieved T$_p$ from the IMACS analysis (1035\,$\pm$\,436\,K) is significantly lower than what has previously been measured for this planet. This low value may result from the use of 1D atmospheric models, see \citet{MacDonald2020}.}. This model has a Bayesian evidence of $\ln{\mathcal{Z}} = 185.1$, marginally preferred over a flat-line model ($\ln{\mathcal{Z}} = 184.6$). The spectral resolution of the IMACS data cannot well-resolve the different TiO and VO absorption features, so either is possible according to our IMACS retrievals. Although not a significant detection of atmospheric molecular absorption ($1.9 \sigma$), the fact that the inclusion of metal oxides - either TiO, VO, or both - raises the evidence of the model hints at a possible consistency between the FORS2, ESPRESSO, and IMACS spectra.


\begin{table}
    \centering
    \caption{Bayesian statistics and model comparison for the retrieval analysis of the FORS2 and IMACS transmission spectra, performed with the {\tt POSEIDON} algorithm. The values after the reference models represent the number of parameters of the model.}
    \begin{tabular}{lccc}
    \hline
         \multirow{2}{*}{Model ($i$)} & Evidence & Detection & Bayes Factor\\
                                & $\ln{\mathcal{Z}_i}$ & of Ref.   & $\ln{\mathcal{B}_{1i}}$ \\
    \hline \hline
        \multicolumn{4}{c}{\textit{FORS2 spectrum} \citep{Sedaghati2017}} \\[2pt]
         Flat (0)           & 803.1     & $\gg 10\,\sigma$  & 173\\
        \multicolumn{4}{c}{\textit{{\tt POSEIDON} atmosphere+activity}} \\[2pt]
         Ref, 16 (1)        & 976.2     & --                & -- \\
         no TiO (2)         & 966.7     & 4.7\,$\sigma$     & 9.5 \\
         no haze (3)        & 965.3     & 5.0\,$\sigma$     & 10.9 \\
         no spots (4)       & 976.6     & NE$^a$            & $-0.4$ \\
        \multicolumn{4}{c}{\textit{{\tt POSEIDON} activity-only}} \\[2pt]
         Spots (5)          & 916.2     & $\gg 10\,\sigma$  & 60 \\
                            &           &                   &  $\ln{\mathcal{B}_{05}}$ = 113 \\

    \hline
        \multicolumn{4}{c}{\textit{IMACS spectrum} \citep{Espinoza2019}} \\[2pt]
         Flat (0)           & 184.6     & NE$^c$           & $-0.3$ \\
                            &           &                   & $\ln{\mathcal{B}_{40}} = 0.5$ \\
        \multicolumn{4}{c}{\textit{{\tt POSEIDON} atmosphere+activity}} \\[2pt]
         Ref, 15 (1)       & 184.3     & --                & -- \\
         no TiO (2)         & 184.0     & 1.5\,$\sigma$     & 0.3 \\
         no TiO+VO (3)      & 183.5     & 1.9\,$\sigma$     & 0.8 \\
         no spots (4)       & 185.1     & BF$^b$            & $-0.8$ \\
    \hline
        \multicolumn{4}{l}{$^a$ NE: No Evidence; where the Bayesian evidence is higher}\\
        \multicolumn{4}{l}{~~~than the reference model.}\\
        \multicolumn{4}{l}{$^b$ BF: Best Fit; for the IMACS data the best fit model}\\
        \multicolumn{4}{l}{~~~is the one with no activity and TiO and/or VO.}\\
        \multicolumn{4}{l}{$^c$ Relative to BF model (4).}
    \end{tabular}
    \label{tab:model statistics}
\end{table}

\section{Summary and Conclusions}
\label{sec:Conclusions}
In this study, we presented high resolution transmission spectroscopy of the hot-Jupiter exoplanet WASP-19b using the cross-dispersed echelle spectrograph ESPRESSO at the ICCF of the VLT. We demonstrated how to mitigate the systematic noise present in the raw frames, originating from the detector readout electronic circuitry and optical path inhomogeneities, as well as a precise correction of telluric absorption signatures in the spectra, using atmospheric profiles measured directly above the observatory as inputs for the models. Possible contamination instances from the laser mounted at a neighbouring telescope were reported, and their impact on narrow-band transmission spectroscopy of the sodium doublet were identified. Stellar and orbital parameters were reported from spectral synthesis analysis and modeling of the Rossiter-McLaughlin effect. Results from both sets of analyses were in agreement with previously reported values, indicating a well-aligned prograde circular orbit for the planet. The impact of center-to-limb variations, as well as the Rossiter-McLaughlin effect, on the calculated transmission spectra (both in the narrow-band and cross correlation analysis) were estimated through modeling of synthetic spectra.

We performed narrow-band transmission spectroscopy for a number of species with significant transition lines. This analysis allowed us to place upper limits on the line contrasts for H, Fe\,{\scriptsize I}, Na\,{\scriptsize I}, Ca\,{\scriptsize I} and K\,{\scriptsize I}. We conducted a cross correlation analysis to search for chemical species, be they atomic or molecular, with a large number of absorption lines and/or bands to systematically combine any transmitted atmospheric signal from their individual lines/bands. Our cross correlation analysis resulted in a non-detection of Fe\,{\scriptsize I}, with an upper limit of $\log\,(X_{\textrm{Fe}}/X_\odot) \approx -1.83\,\pm\,0.11$, namely sub-solar abundances. We found that the observed ESPRESSO data was insufficient to detect the presence of H$_2$O in the upper atmosphere for previously retrieved abundances, due to the lack of prominent absorption bands over the ESPRESSO wavelength range, and consequently it was not possible to confirm or reject previous detections of H$_2$O at low spectral resolution \citep{Sing2016,Sedaghati2017}. We detected a 3.02\,$\pm$\,0.15\,$\sigma$ peak for TiO models in the CCF map, located at the radial velocity of the planet and the systemic velocity of the star. This peak is consistent with sub-solar TiO abundances injected into the data, in agreement with previously retrieved values. However, we refrain from claiming confirmation of the previously reported detection of TiO from low resolution FORS2 transmission spectra \citep{Sedaghati2017}, as typically detections from cross correlation analysis require higher levels of significance. We again detect the presence of a strong scattering towards blue wavelengths from chromatic Rossiter-McLaughlin measurements, consistent with the slope detected by \citet{Sedaghati2017} and in contrast with the IMACS results \citep{Espinoza2019}.

Finally, we discussed the implications of including unocculted stellar active regions in the atmospheric retrieval analyses of existing low resolution transmission spectra of WASP-19b. We fitted two of the individual data sets comprising the FORS2 spectrum with an activity-only model, showing that the presence of stellar spots covering large portions of the stellar disk ($\sim41 - 57\%$) could explain some of the enhanced planetary radius towards the blue wavelengths and possibly some of the variations attributed to TiO molecular absorption. However, such models require retrieved stellar photosphere temperatures of $\sim$\,$3760 - 4250$\,K, far and significantly below previously measured values. We further showed that a spot-only retrieval of the full FORS2 transmission spectrum is an overall poor fit to the observations. Therefore, we conclude that stellar activity alone is not responsible for the observed features in the FORS2 transmission spectrum. We additionally performed retrievals considering the simultaneous presence of unocculted active regions and a planetary atmosphere, finding atmospheric properties for WASP-19b largely consistent with \citet{Sedaghati2017}. While these retrievals allow for the presence of a population of spots $\sim$1200\,K below the surrounding photosphere covering $\sim$10\,\% of the stellar surface, Bayesian model comparisons do not statistically favour the inclusion of unocculted spots. Our model comparisons establish a revised 4.7\,$\sigma$ detection of atmospheric TiO (factoring in the potential influence of spots) and 5.0\,$\sigma$ detection of scattering due to hazes towards blue wavelengths. We presented an updated, precise, TiO abundance for WASP-19b: $\log n_{\textrm{{\tiny TiO}}} = -9.02 \pm 0.45$. This $\sim$\,100$\times$ sub-solar TiO abundance, coupled with our non-detection of Fe from ESPRESSO, could constitute direct observational evidence that cold trapping processes are depleting heavy metal abundances at WASP-19b's terminator \citep[e.g.][]{Parmentier2013}.


To reconcile differing atmospheric inferences from the WASP-19b FORS2 analysis of \citet{Sedaghati2017} and the IMACS analysis of \citet{Espinoza2019}, we also performed atmospheric retrievals on the IMACS dataset. We conclude that this data, while relatively featureless, is best explained by a model including absorption from the metals oxides TiO and/or VO. The inclusion of these atmospheric molecules results in a marginal improvement over flat-line or activity-only models. We therefore suggest that the IMACS data are indeed consistent with the presence of TiO in WASP-19b's atmosphere, despite the disagreement in spectral morphology between the IMACS and FORS2 data - the latter being consistent with our ESPRESSO-derived chromatic Rossiter-McLaughlin spectrum.


This study provided further evidence suggesting the presence of TiO in the upper atmosphere of WASP-19b, and presented a pathway to reconciling previous results from low resolution transmission spectra. Given current instrumentation, a further avenue for probing the atmosphere is to observe the short primary transit with ESPRESSO in 4-UT mode, with the goal of increasing S/N in the CCF map. WASP-19b remains an intriguing target for atmospheric studies, holding the tantalising prospect of revealing the physics and chemistry of heavy metals in hot-Jupiter atmospheres.

\section*{Acknowledgements}
We would like to thank the anonymous referee for the constructive revision of the manuscript. Addressing the points raised certainly improved the quality of the presented work significantly. We acknowledge the use of the following {\tt python} packages in addition to those explicitly mentioned in the text: {\tt NumPy} \citep{NumPy}, {\tt matplotlib} \citep{matplotlib}, {\tt SciPy} \citep{SciPy}, {\tt pandas} \citep{pandas}, {\tt emcee} \citep{emcee}, {\tt scikit-learn} \citep{scikit}, {\tt numba} \citep{numba}, {\tt corner} \citep{corner} and {\tt astropy} \citep{astropy}. This work has made use of the {\tt VALD} database, operated at Uppsala University, the Institute of Astronomy RAS in Moscow, and the University of Vienna. ES\ acknowledges support from ESO through its fellowship program.
RB\ acknowledges support from FONDECYT Project 11200751, and CORFO project N$^\circ$14ENI2-26865. RB acknowledges support from ANID -- Millennium Science Initiative -- ICN12\_009.

\section*{Data availability}
This work has been made possible through observations performed at ESO's Paranal observatory, the data for which are publicly available from the ESO data archives (\url{http://archive.eso.org/eso/eso_archive_main.html}), under the program ID 0102.C-0311. The reduced data are also available upon request from E. Sedaghati. 




\bibliographystyle{mnras}
\bibliography{bibliography} 


\appendix
\section{RV values}
\label{appendix:RV data}
In this section we present the RV values calculated by the ESPRESSO DRS, which are used in modeling the RM effect in section \ref{sec:Star model}.

\begin{table}
    \centering
    \begin{tabular}{cccc}
    \hline
    BJD\,[days] & RV\,[km/s] & $\sigma_{\textrm{RV}}$\,[m/s] & EXP\,[s]\\
    \hline \hline
2458498.58988800  & 20.82646  & 8.18  & 930 \\
2458498.60100021  & 20.7993   & 7.16  & 930 \\
2458498.61199321  & 20.7785   & 8.19  & 930 \\
2458498.62379173  & 20.76231  & 6.51  & 930 \\
2458498.63374199  & 20.76321  & 14.5  & 380 \\
2458498.63892399  & 20.77791  & 12.74 & 380 \\
2458498.64384147  & 20.79349  & 13.38 & 380 \\
2458498.64876042  & 20.73967  & 14.24 & 380 \\
2458498.65356825  & 20.69906  & 14.62 & 380 \\
2458498.65861991  & 20.69273  & 14.02 & 380 \\
2458498.66350586  & 20.67341  & 14.75 & 380 \\
2458498.66846226  & 20.63671  & 13.24 & 380 \\
2458498.67348508  & 20.61285  & 10.11 & 380 \\
2458498.67826987  & 20.60608  & 8.99  & 380 \\
2458498.68327102  & 20.61570  & 8.97  & 380 \\
2458498.68819888  & 20.61782  & 7.81  & 380 \\
2458498.69307304  & 20.62464  & 7.45  & 380 \\
2458498.70007611  & 20.61316  & 4.57  & 760 \\
2458498.70964022  & 20.60040  & 4.56  & 760 \\
2458498.71876645  & 20.59894  & 4.35  & 760 \\
2458498.72840475  & 20.56458  & 4.39  & 760 \\
2458498.73782029  & 20.55429  & 3.99  & 800 \\
    \hline
    \end{tabular}
    \caption{The pipeline calculated RV values for observations in DS1, as well as their uncertainties and exposure times used.}
    \label{tab:RV1}
\end{table}

\begin{table}
    \centering
    \begin{tabular}{cccc}
    \hline
    BJD\,[days] & RV\,[km/s] & $\sigma_{\textrm{RV}}$\,[m/s] & EXP\,[s]\\
    \hline \hline
2458546.70029843 & 20.81404 & 5.42  & 1200 \\
2458546.71730722 & 20.77644 & 8.64  & 1200 \\
2458546.73022202 & 20.74880 & 9.11  & 1200 \\
2458546.74635059 & 20.74357 & 7.49  & 1200 \\
2458546.75597151 & 20.74175 & 17.71 & 410  \\
2458546.76164054 & 20.70465 & 18.3  & 410  \\
2458546.76693892 & 20.74182 & 20.89 & 410  \\
2458546.77253972 & 20.66661 & 15.99 & 410  \\
2458546.77779036 & 20.65585 & 15.02 & 410  \\
2458546.78315677 & 20.66390 & 15.14 & 410  \\
2458546.78829173 & 20.57827 & 20.02 & 410  \\
2458546.79395184 & 20.59739 & 18.90 & 410  \\
2458546.79956324 & 20.55623 & 20.08 & 410  \\
2458546.80473385 & 20.63033 & 17.92 & 410  \\
2458546.81027239 & 20.62332 & 17.76 & 410  \\
2458546.81558596 & 20.59751 & 16.65 & 410  \\
2458546.82247144 & 20.61843 & 15.00 & 670  \\
2458546.83106347 & 20.50721 & 13.37 & 670  \\
2458546.83938853 & 20.57332 & 11.28 & 670  \\
2458546.84775486 & 20.55209 & 15.56 & 670  \\
2458546.85650149 & 20.51568 & 13.58 & 670  \\
    \hline
    \end{tabular}
    \caption{Same as Table \ref{tab:RV1} but for DS2.}
    \label{tab:RV2}
\end{table}

\begin{table}
    \centering
    \begin{tabular}{cccc}
    \hline
    BJD\,[days] & RV\,[km/s] & $\sigma_{\textrm{RV}}$\,[m/s] & EXP\,[s]\\
    \hline \hline
2458565.62926856 & 20.85050 & 4.19 & 830 \\
2458565.63970318 & 20.83263 & 4.15 & 830 \\
2458565.64988114 & 20.81295 & 3.92 & 830 \\
2458565.66009538 & 20.79194 & 4.31 & 830 \\
2458565.67041463 & 20.76472 & 4.20 & 830 \\
2458565.68037855 & 20.77662 & 3.64 & 830 \\
2458565.68812129 & 20.78292 & 8.50 & 380 \\
2458565.69318121 & 20.76429 & 8.68 & 380 \\
2458565.69819600 & 20.74261 & 8.36 & 380 \\
2458565.70319448 & 20.72989 & 8.13 & 380 \\
2458565.70815185 & 20.68067 & 8.44 & 380 \\
2458565.71324232 & 20.66203 & 8.96 & 380 \\
2458565.71823768 & 20.64335 & 8.30 & 380 \\
2458565.72317414 & 20.61127 & 8.58 & 380 \\
2458565.72822643 & 20.61737 & 8.65 & 380 \\
2458565.73323846 & 20.61288 & 7.69 & 380 \\
2458565.73821655 & 20.62037 & 7.40 & 380 \\
2458565.74326025 & 20.61360 & 7.33 & 380 \\
2458565.75020172 & 20.60935 & 3.91 & 720 \\
2458565.75912765 & 20.59612 & 3.89 & 720 \\
2458565.76806310 & 20.57339 & 3.99 & 720 \\
2458565.77696335 & 20.56269 & 4.03 & 720 \\
2458565.78279489 & 20.54291 & 15.37 & 180.0\\
    \hline
    \end{tabular}
    \caption{Same as Table \ref{tab:RV1} but for DS3.}
    \label{tab:RV3}
\end{table}

\begin{table}
    \centering
    \begin{tabular}{cccc}
    \hline
    BJD\,[days] & RV\,[km/s] & $\sigma_{\textrm{RV}}$\,[m/s] & EXP\,[s]\\
    \hline \hline
2458860.65788891 & 20.80633 & 2.47 & 900 \\
2458860.66873710 & 20.78622 & 2.19 & 900 \\
2458860.68006353 & 20.77083 & 2.24 & 900 \\
2458860.69152885 & 20.75252 & 2.05 & 900 \\
2458860.70249466 & 20.73853 & 2.01 & 900 \\
2458860.71155150 & 20.76021 & 3.06 & 380 \\
2458860.71684916 & 20.75988 & 3.29 & 380 \\
2458860.72216158 & 20.74915 & 2.98 & 380 \\
2458860.72762300 & 20.72062 & 2.96 & 380 \\
2458860.73294862 & 20.68917 & 3.05 & 380 \\
2458860.73822609 & 20.65960 & 2.89 & 380 \\
2458860.74354597 & 20.62669 & 2.89 & 380 \\
2458860.74880799 & 20.60702 & 3.00 & 380 \\
2458860.75419957 & 20.59131 & 2.93 & 380 \\
2458860.75953740 & 20.59485 & 2.85 & 380 \\
2458860.76491886 & 20.60629 & 2.80 & 380 \\
2458860.77016483 & 20.60589 & 2.88 & 380 \\
2458860.77767389 & 20.59486 & 1.95 & 750 \\
2458860.78728462 & 20.57681 & 2.00 & 750 \\
2458860.79690229 & 20.56175 & 2.08 & 750 \\
2458860.80650307 & 20.54011 & 2.13 & 750 \\
2458860.81598265 & 20.52403 & 2.11 & 750 \\
    \hline
    \end{tabular}
    \caption{Same as Table \ref{tab:RV1} but for DS4.}
    \label{tab:RV4}
\end{table}

\section{RM fit posteriors}

\begin{figure}
	\includegraphics[width=\columnwidth]{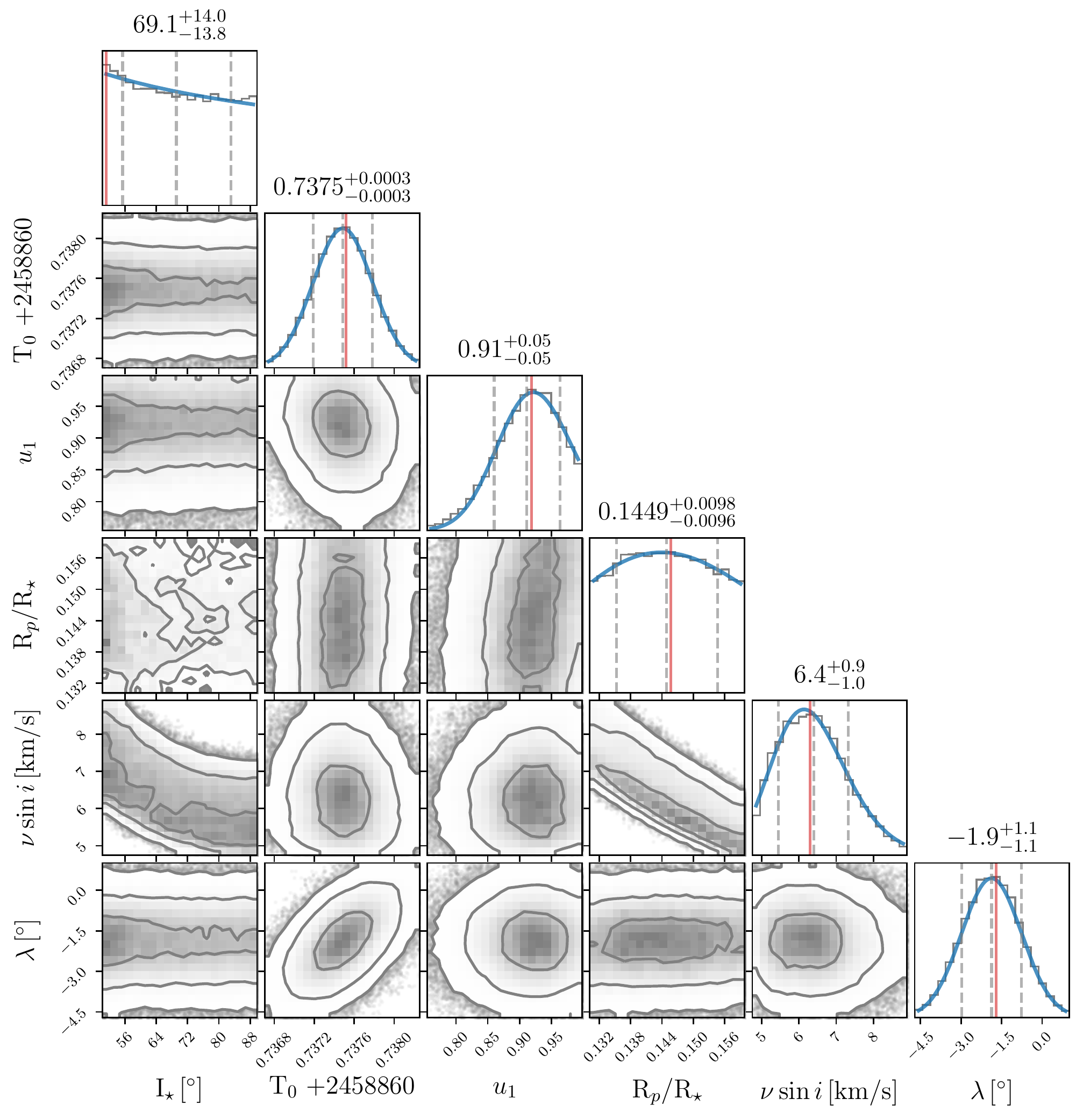}
    \caption{Posterior probability distributions for the parameters of the RM model fit to the RV variations measured for DS4. The vertical gray dashed lines in the diagonal plots indicate the mean and $\pm 1\sigma$ levels of the distributions and the red line shows the maximum of the fitted log-normal models (blue curves).}
    \label{fig:RM4 posteriors}
\end{figure}

\section{CRM RV values}
\newpage
\begin{table*}
    \centering
    \caption{Radial velocity values calculated for the purpose of CRM analysis from DS4. Column headers indicate the centres of the wavelength bins, each created from 5 spectra orders. The data is plotted in Fig. \ref{fig:Chromatic RM}, where the time stamps are the same as those given in table \ref{tab:RV4}.}
    \label{tab:CRM RVs}
    \begin{tabular}{cccccccc}
    \hline
    3852\,\AA & 3977\,\AA & 4111\,\AA & 4254\,\AA & 4407\,\AA & 4571\,\AA & 4748\,\AA & 4940\,\AA \\
    $[$km/s$]$ & [km/s] & [km/s] & [km/s] & [km/s] & [km/s] & [km/s] & [km/s] \\ 
    \hline \hline
$21.160 \pm 0.041$ & $21.013 \pm 0.017$ & $20.949 \pm 0.008$ & $20.960 \pm 0.008$ & $21.079 \pm 0.010$ & $21.084 \pm 0.009$ & $21.004 \pm 0.006$ & $21.025 \pm 0.006$ \\
$21.114 \pm 0.041$ & $20.968 \pm 0.015$ & $20.928 \pm 0.012$ & $20.946 \pm 0.006$ & $21.049 \pm 0.011$ & $21.075 \pm 0.012$ & $21.000 \pm 0.006$ & $20.999 \pm 0.004$ \\
$21.211 \pm 0.022$ & $20.951 \pm 0.013$ & $20.909 \pm 0.008$ & $20.918 \pm 0.008$ & $21.019 \pm 0.010$ & $21.067 \pm 0.010$ & $20.979 \pm 0.008$ & $20.982 \pm 0.006$ \\
$21.091 \pm 0.030$ & $20.914 \pm 0.014$ & $20.882 \pm 0.007$ & $20.902 \pm 0.007$ & $21.026 \pm 0.010$ & $21.048 \pm 0.012$ & $20.953 \pm 0.008$ & $20.970 \pm 0.005$ \\
$21.030 \pm 0.033$ & $20.915 \pm 0.010$ & $20.882 \pm 0.007$ & $20.897 \pm 0.006$ & $21.006 \pm 0.011$ & $21.026 \pm 0.010$ & $20.936 \pm 0.007$ & $20.950 \pm 0.006$ \\
$21.166 \pm 0.030$ & $20.947 \pm 0.023$ & $20.881 \pm 0.014$ & $20.915 \pm 0.009$ & $21.012 \pm 0.012$ & $21.058 \pm 0.013$ & $20.962 \pm 0.007$ & $20.970 \pm 0.007$ \\
$21.074 \pm 0.045$ & $20.945 \pm 0.025$ & $20.905 \pm 0.014$ & $20.914 \pm 0.011$ & $21.022 \pm 0.009$ & $21.050 \pm 0.016$ & $20.959 \pm 0.009$ & $20.973 \pm 0.007$ \\
$21.169 \pm 0.048$ & $20.948 \pm 0.018$ & $20.871 \pm 0.014$ & $20.951 \pm 0.009$ & $21.001 \pm 0.011$ & $21.029 \pm 0.011$ & $20.931 \pm 0.007$ & $20.960 \pm 0.008$ \\
$21.069 \pm 0.047$ & $20.905 \pm 0.013$ & $20.858 \pm 0.011$ & $20.870 \pm 0.008$ & $20.996 \pm 0.010$ & $21.021 \pm 0.009$ & $20.921 \pm 0.008$ & $20.944 \pm 0.008$ \\
$21.089 \pm 0.053$ & $20.886 \pm 0.022$ & $20.831 \pm 0.013$ & $20.824 \pm 0.009$ & $20.945 \pm 0.010$ & $20.991 \pm 0.011$ & $20.875 \pm 0.007$ & $20.915 \pm 0.007$ \\
$21.044 \pm 0.062$ & $20.822 \pm 0.020$ & $20.793 \pm 0.011$ & $20.813 \pm 0.009$ & $20.926 \pm 0.013$ & $20.957 \pm 0.014$ & $20.855 \pm 0.008$ & $20.884 \pm 0.006$ \\
$20.977 \pm 0.036$ & $20.775 \pm 0.021$ & $20.756 \pm 0.012$ & $20.773 \pm 0.012$ & $20.891 \pm 0.013$ & $20.928 \pm 0.014$ & $20.810 \pm 0.007$ & $20.845 \pm 0.008$ \\
$20.943 \pm 0.049$ & $20.818 \pm 0.023$ & $20.749 \pm 0.014$ & $20.782 \pm 0.010$ & $20.866 \pm 0.014$ & $20.889 \pm 0.013$ & $20.792 \pm 0.008$ & $20.814 \pm 0.007$ \\
$21.002 \pm 0.053$ & $20.823 \pm 0.022$ & $20.714 \pm 0.012$ & $20.768 \pm 0.008$ & $20.845 \pm 0.011$ & $20.873 \pm 0.008$ & $20.768 \pm 0.008$ & $20.812 \pm 0.007$ \\
$20.967 \pm 0.044$ & $20.785 \pm 0.021$ & $20.752 \pm 0.014$ & $20.738 \pm 0.011$ & $20.874 \pm 0.011$ & $20.894 \pm 0.013$ & $20.795 \pm 0.006$ & $20.813 \pm 0.007$ \\
$20.917 \pm 0.052$ & $20.810 \pm 0.021$ & $20.748 \pm 0.013$ & $20.768 \pm 0.010$ & $20.896 \pm 0.013$ & $20.889 \pm 0.013$ & $20.793 \pm 0.008$ & $20.812 \pm 0.004$ \\
$20.968 \pm 0.050$ & $20.762 \pm 0.016$ & $20.743 \pm 0.010$ & $20.757 \pm 0.009$ & $20.878 \pm 0.010$ & $20.906 \pm 0.008$ & $20.807 \pm 0.006$ & $20.825 \pm 0.007$ \\
$20.930 \pm 0.037$ & $20.801 \pm 0.015$ & $20.738 \pm 0.009$ & $20.765 \pm 0.006$ & $20.864 \pm 0.010$ & $20.877 \pm 0.009$ & $20.788 \pm 0.004$ & $20.810 \pm 0.006$ \\
$20.871 \pm 0.028$ & $20.759 \pm 0.014$ & $20.717 \pm 0.008$ & $20.742 \pm 0.007$ & $20.852 \pm 0.011$ & $20.861 \pm 0.010$ & $20.777 \pm 0.006$ & $20.804 \pm 0.006$ \\
$20.921 \pm 0.027$ & $20.758 \pm 0.014$ & $20.706 \pm 0.008$ & $20.737 \pm 0.006$ & $20.831 \pm 0.010$ & $20.848 \pm 0.010$ & $20.769 \pm 0.006$ & $20.775 \pm 0.004$ \\
$20.928 \pm 0.029$ & $20.716 \pm 0.015$ & $20.682 \pm 0.011$ & $20.702 \pm 0.008$ & $20.789 \pm 0.010$ & $20.823 \pm 0.011$ & $20.745 \pm 0.005$ & $20.750 \pm 0.006$ \\
$20.839 \pm 0.036$ & $20.701 \pm 0.014$ & $20.679 \pm 0.011$ & $20.687 \pm 0.009$ & $20.793 \pm 0.011$ & $20.807 \pm 0.012$ & $20.722 \pm 0.006$ & $20.741 \pm 0.005$ \\
\hline
5148\,\AA & 5329\,\AA & 5571\,\AA & 5848\,\AA & 6190\,\AA & 6563\,\AA & 7389\,\AA & \\
$[$km/s$]$ & [km/s] & [km/s] & [km/s] & [km/s] & [km/s] & [km/s] & \\
\hline \hline
$20.943 \pm 0.007$ & $21.013 \pm 0.007$ & $20.960 \pm 0.010$ & $21.006 \pm 0.007$ & $21.026 \pm 0.009$ & $20.960 \pm 0.008$ & $21.096 \pm 0.015$ & \\
$20.925 \pm 0.007$ & $20.993 \pm 0.007$ & $20.937 \pm 0.026$ & $20.967 \pm 0.010$ & $21.022 \pm 0.007$ & $20.951 \pm 0.009$ & $21.101 \pm 0.007$ & \\
$20.915 \pm 0.006$ & $20.980 \pm 0.006$ & $20.925 \pm 0.009$ & $20.968 \pm 0.007$ & $21.014 \pm 0.008$ & $20.937 \pm 0.010$ & $21.047 \pm 0.008$ & \\
$20.906 \pm 0.006$ & $20.965 \pm 0.005$ & $20.908 \pm 0.008$ & $20.948 \pm 0.009$ & $20.969 \pm 0.007$ & $20.925 \pm 0.010$ & $21.048 \pm 0.009$ & \\
$20.877 \pm 0.006$ & $20.961 \pm 0.006$ & $20.904 \pm 0.007$ & $20.925 \pm 0.010$ & $20.966 \pm 0.007$ & $20.907 \pm 0.008$ & $21.026 \pm 0.011$ & \\
$20.896 \pm 0.011$ & $20.978 \pm 0.010$ & $20.936 \pm 0.009$ & $20.957 \pm 0.013$ & $20.990 \pm 0.009$ & $20.930 \pm 0.011$ & $21.062 \pm 0.012$ & \\
$20.897 \pm 0.008$ & $20.991 \pm 0.011$ & $20.911 \pm 0.009$ & $20.930 \pm 0.013$ & $20.984 \pm 0.010$ & $20.933 \pm 0.016$ & $21.085 \pm 0.016$ & \\
$20.893 \pm 0.008$ & $20.970 \pm 0.010$ & $20.914 \pm 0.013$ & $20.949 \pm 0.012$ & $20.966 \pm 0.008$ & $20.908 \pm 0.012$ & $21.032 \pm 0.013$ & \\
$20.855 \pm 0.007$ & $20.940 \pm 0.009$ & $20.874 \pm 0.010$ & $20.899 \pm 0.012$ & $20.936 \pm 0.007$ & $20.883 \pm 0.012$ & $21.009 \pm 0.015$ & \\
$20.829 \pm 0.007$ & $20.899 \pm 0.008$ & $20.852 \pm 0.009$ & $20.905 \pm 0.012$ & $20.918 \pm 0.008$ & $20.837 \pm 0.016$ & $20.972 \pm 0.012$ & \\
$20.800 \pm 0.006$ & $20.849 \pm 0.011$ & $20.805 \pm 0.008$ & $20.827 \pm 0.010$ & $20.886 \pm 0.010$ & $20.785 \pm 0.011$ & $20.931 \pm 0.016$ & \\
$20.746 \pm 0.007$ & $20.844 \pm 0.008$ & $20.780 \pm 0.009$ & $20.798 \pm 0.012$ & $20.874 \pm 0.010$ & $20.821 \pm 0.010$ & $20.923 \pm 0.016$ & \\
$20.729 \pm 0.008$ & $20.820 \pm 0.009$ & $20.753 \pm 0.008$ & $20.776 \pm 0.013$ & $20.819 \pm 0.007$ & $20.744 \pm 0.016$ & $20.888 \pm 0.010$ & \\
$20.731 \pm 0.007$ & $20.788 \pm 0.008$ & $20.727 \pm 0.009$ & $20.759 \pm 0.010$ & $20.806 \pm 0.010$ & $20.750 \pm 0.016$ & $20.874 \pm 0.012$ & \\
$20.712 \pm 0.007$ & $20.810 \pm 0.011$ & $20.750 \pm 0.012$ & $20.791 \pm 0.013$ & $20.824 \pm 0.007$ & $20.742 \pm 0.011$ & $20.862 \pm 0.014$ & \\
$20.746 \pm 0.006$ & $20.812 \pm 0.009$ & $20.764 \pm 0.010$ & $20.792 \pm 0.009$ & $20.820 \pm 0.010$ & $20.780 \pm 0.011$ & $20.912 \pm 0.016$ & \\
$20.749 \pm 0.007$ & $20.822 \pm 0.010$ & $20.760 \pm 0.010$ & $20.802 \pm 0.010$ & $20.799 \pm 0.010$ & $20.784 \pm 0.013$ & $20.888 \pm 0.014$ & \\
$20.732 \pm 0.007$ & $20.806 \pm 0.008$ & $20.745 \pm 0.008$ & $20.777 \pm 0.008$ & $20.814 \pm 0.007$ & $20.760 \pm 0.010$ & $20.846 \pm 0.012$ & \\
$20.712 \pm 0.006$ & $20.788 \pm 0.007$ & $20.721 \pm 0.008$ & $20.762 \pm 0.010$ & $20.803 \pm 0.008$ & $20.739 \pm 0.008$ & $20.881 \pm 0.012$ & \\
$20.691 \pm 0.008$ & $20.756 \pm 0.008$ & $20.726 \pm 0.007$ & $20.745 \pm 0.011$ & $20.795 \pm 0.006$ & $20.721 \pm 0.014$ & $20.879 \pm 0.010$ & \\
$20.678 \pm 0.005$ & $20.751 \pm 0.007$ & $20.700 \pm 0.008$ & $20.739 \pm 0.006$ & $20.755 \pm 0.005$ & $20.714 \pm 0.010$ & $20.818 \pm 0.015$ & \\
$20.659 \pm 0.005$ & $20.730 \pm 0.008$ & $20.688 \pm 0.008$ & $20.711 \pm 0.009$ & $20.759 \pm 0.007$ & $20.691 \pm 0.010$ & $20.845 \pm 0.011$ & \\
\hline
    \end{tabular}
\end{table*}


\bsp	
\label{lastpage}
\end{document}